\documentclass[10pt]{article}
\usepackage{CJK}   
\usepackage{indentfirst}  
\usepackage{latexsym}
\usepackage{amsmath}
\usepackage{amssymb}
\usepackage{bm}
\usepackage{graphicx}
\usepackage{epstopdf}
\usepackage{epsfig}
\usepackage{multirow}
\usepackage{cite}

\def\be{\begin{eqnarray}}
\def\ee{\end{eqnarray}}

\def\({\left(}
\def\){\right)}
\def\bc{\begin{center}}

\def\l{\label}

\def\ec{\end{center}}
\def\no{\nonumber}

\def\bey{\begin{eqnarray*}}
\def\eey{\end{eqnarray*}}
\def\ber{\begin{array}{l}}
\def\eer{\end{array}}

\newcommand{\h}{\hspace{0.06cm}}
\newtheorem{thh}{Theorem}

\newtheorem{lm}{Lemma}

\usepackage{bm}
\begin{document}
\date{}
\title{
The finite dimensional subalgebra classification of infinite dimensional symmetry algebra of two dimensional coupled  nonlinear Schr\"{o}dinger equations\footnote{ Supported by National Natural Science Foundation of China (11571008).}\\}
\author{\small YueXing Bai$^1$,Temuer Chaolu$^2$\footnote{\small Corresponding author E-mail: tmchaolu@shmtu.edu.cn}, Yan Li$^2$\\
        {\small (1.College of Arts and Sciences, Shanghai Maritime University, Shanghai 200135, China)}\\
        {\small (2. Inner Mongolia University of Technology, Huhhot,010051, China)}}
\maketitle

{\bf Abstract: }The symmetry group structures of two dimensional coupled nonlinear Shr\"{o}dinger equations are considered. We first show that the equations admit infinite dimensional symmetry algebra as well as the corresponding symmetry group depending on four arbitrary functions of one variable. Then we show some physical symmetries and an affine loop algebra contained in the symmetry algebra of the equations. Third, we give the complete classifications of finite dimension (less than four) subalgebras of the symmetry algebra under the adjoint group of the symmetry group. These results provide the theoretical and computational basis for the further study of the equations with symmetry methods.


\section{Introduction}
The nonlinear Shr\"{o}dinger typed equations are of considerable importance in both theory and applications. Particularly, the various coupled nonlinear Schr\"{o}dinger systems are used in different fields of physics and mathematics \cite{A0}. One of them, a system of (2+1)-dimensional coupled nonlinear Schr\"{o}dinger (2D-CNLS) system
\begin{equation}
\Omega:\,\,\, \left\{
\begin{array}{l}
iu_{t}-pu_{xx}+qu_{yy}+r|u|^{2}u-2uv=0,\\
pv_{xx}+qv_{yy}-pr(|u|^{2})_{xx}=0,\label{e1}
\end{array}
\right.
\end{equation}
arising in the describing the evolution of the wave packet on a two-dimensional water surface under gravity\cite{c1, c2},  has been investigated  in series of papers \cite{c3, c4, c5, c6, c7, c8, c9, c10, Ali}. The evolutions of packets of various waves were discussed in \cite{c3,c4,c5,c6}; the multi-soliton solutions were obtained in \cite{c7} in which the two-soliton resonant interaction and a triple soliton structure were found; different types of lump solutions were found in \cite{c9, c8}; particular symmetry reductions and corresponding similarity solutions are obtained in {\cite{c10}}. The one dimensional optimal system of an eight dimensional symmetry algebra and corresponding reductions of system (\ref{e1}) are given in \cite{Ali}.

Actually, as we showed in \cite{Ali}, the system (\ref{e1}) admits infinite dimensional symmetry (Lie) algebra of the system (we call it as 2D-CNLS algebra) depending on four arbitrary functions of single variable $t$ with corresponding infinite dimensional symmetry group (we call it as 2D-CNLS group). As usual, to solve system (\ref{e1}) by the symmetry method, we need to identify  the  subalgebras of the 2D-CNLS algebra. More particularly, we need subalgebras that correspond to Lie groups having orbits of codimension less than four \cite{Ib}. In general sense, what structures of finite dimensional subalgebras does an infinite dimensional algebra have and how to identify them are nontrivial. This  problem is equivalent to the subalgebra classification problem under an algebraic equivalent criterion. For finite dimensional Lie algebra, the classification methods have been developed in \cite{Ib, class, patera, B2}. In \cite{main1, main2}, the authors applying the methods gave the subalgebra classification of infinite dimensional symmetry algebras of the Kadomtsev- Petviashvili equation and the selfdual Einstein equations. These are the pioneering works of investigating the structure of infinite dimensional symmetry group (algebra) admitted by a partial differential equations (PDEs). It has been turned out that many of the important nonlinear PDEs of modern physics admit infinite dimensional symmetry algebra depending on arbitrary functions. Such are the cases studied in \cite{main1, main2} and 2D-CNLS treated in this article, but also of other PDEs in higher dimensions. For instance, in Earth science, the PDEs of dynamic convection in a sea \cite{I1} admit infinite dimensional symmetry algebra with five arbitrary functions; In incompressible fluids, the two (three) dimensional Navier- Stokes equations admit symmetry algebra with three (four) arbitrary functions \cite{7}; in nonlinear optics and acoustics, the transparent atmosphere equations admit symmetry algebra with five arbitrary functions \cite{ku}; in geophysical fluid dynamics the two dimensional nonlinear Boussinesq equations admit symmetry algebra with three arbitrary functions \cite{I2}. The presence of arbitrary functions in the symmetry algebra is a characteristic property of such science fields \cite{7, 2}. However the structure of the finite dimensional subalgebras of the such infinite dimensional symmetry algebra of a PDEs and the corresponding complete reductions of the PDEs  have yet not been fully studied.

In this article, using the methods in \cite{main1}, we give the classification of the subalgebras of dimension less than four of the 2D-CNLS algebra under the adjoint group (action) of the 2D-CNLS group (the group of inner automorphisms of the 2D-CNLS algebra). It will derive a better understanding of the structure of the 2D-CNLS algebra and clarify the invariant solutions of different classes of the system (\ref{e1}) under the adjoint group. The classification also further show the applicability of the methods developed for classifying optimal subalgebras of finite-dimensional Lie algebras.

The present paper is organized as follows: In Sec. II, we show the 2D-CNLS algebra and the corresponding 2D-CNLS group. We establish a Levi decomposition of the algebra. For Laurent polynomials form of the arbitrary functions in the algebra, we identify the Lie algebra as a subalgebra of the loop algebra $A_4^{(1)}$, a Kac- Moody type of algebra. We also present the physical meaning of the finite-dimensional algebras obtained by restricting the arbitrary functions involved in 2D-CNLS algebra to first and second degree polynomials. In Sec. III, the properties of the action of the adjoint group corresponding to the basis elements of the 2D-CNLS algebra are presented. These properties are frequently used in the computations for obtaining the classification of the 2D-CNLS algebra. In Sec. IV, as the heart of present article, the classification of the less then four dimension subalgebras under the adjoint action is presented.  Finally, in Sec.V, we give some brief conclusion remarks on our results.

Here  a remark about some notations frequently used in this article.

\textbf{Remark 1: }

Use $\mathbb{R}$ denote the complex number and real number fields respectively.

Notation $C^\infty(\mathbb{R})$ represents the set of infinite order differential functions on $\mathbb{R}$.

Use notations $f, g, h, k, \phi,\psi,\Phi, \Psi, F, G, K, H$ to denote the arbitrary functions of single variable $t$ in $C^\infty (\mathbb{R})$ and use $\alpha, \beta, \lambda, \delta, \mu, \nu, \sigma, \gamma, \rho, \kappa, a, b, c, d$ to denote arbitrary constants in $\mathbb{R}$.

Denote $\epsilon=q/p \in \mathbb{R}$ for physical constants $p$ and $q$ in 2D-CNLS system (\ref{e1}).

\section{The 2D-CNLS algebra and group}
\subsection{The 2D-CNLS algebra}

Let the 2D-CNLS group, i.e. an one-parameter Lie's symmetry group of infinitesimal transformation of system (\ref{e1}) mapping  $(x,y,t,u,v)$ into $(x^\prime, y^\prime, t^\prime, u^\prime, v^\prime)$, be given as\cite{B1,B2}
\begin{equation}
\left\{
\begin{array}{l}
x^\prime=x+\varepsilon \xi(x,y,t,u,v)+O(\varepsilon^{2}),\\
y^\prime=y+\varepsilon \eta(x,y,t,u,v)+O(\varepsilon^{2}),\\
t^\prime=t+\varepsilon \tau(x,y,t,u,v)+O(\varepsilon^{2}),\\
u^\prime=u+\varepsilon \zeta(x,y,t,u,v)+O(\varepsilon^{2}),\\
v^\prime=v+\varepsilon \varsigma(x,y,t,u,v)+O(\varepsilon^{2}).\label{e2}
\end{array}
\right.
\end{equation}

Thus the symmetry group is generated by the infinitesimal generator
\begin{eqnarray}
\mathcal{X}=\xi\frac{\partial}{\partial x}+\eta\frac{\partial}{\partial y}+\tau\frac{\partial}{\partial t}+\zeta\frac{\partial}{\partial u}
+\varsigma\frac{\partial}{\partial v},\l{e3}
\end{eqnarray}
where $\xi=\xi(x,y,t,u,v)$, $\eta=\eta(x,y,t,u,v)$, $\tau=\tau (x,y,t,u,v)$,
$\zeta=\zeta (x,y,t,u,v)$ and $\varsigma=\varsigma(x,y,t,u,v)$
are the infinitesimal functions of $\mathcal{X}$ belonging to $C^\infty(\mathbb{R}^5)$.  The invariant condition of equation (\ref{e1}) under (\ref{e2}) is given by
$
\textrm{Pr}^{(2)}\mathcal{X}( \Omega)|_{\Omega=0}=0,
$
where {Pr}$^{(2)}\mathcal{X}$ is the second prolongation of generator $\mathcal{X}$ on the jut space of $u$ and $v$. Algorithms for determining the generator (\ref{V}) of symmetry groups of a PDEs are well known \cite{B1, B2} and also see recent new algorithm in \cite{chao5, B7}. For the 2D-CNLS system (\ref{e1}), a straightforward application of the algorithm has yielded an infinite dimensional Lie algebra of the symmetry group with the infinitesimal functions given by
\begin{eqnarray*}
&&\xi=f(t)+\frac{1}{2}x\dot{h }(t),\,\, \eta=g(t)+\frac{1}{2}y \dot{h}(t),\,\, \tau=h(t), \nonumber\\
&&\zeta=\left(ik(t)-\frac{1}{2}\dot{h}(t)-\frac{i}{8pq}(4(q x  \dot{f}(t)-p y  \dot{g}(t))+(q x^{2}-p y^{2}) \ddot{h}(t)\right)u, \nonumber\\
&&\varsigma=-\dot{h}(t)v -\frac{1}{2}\dot{k}(t)-\frac{1}{16pq}\left(8 (q x\ddot{f}(t)-p y \ddot{g}(t))+(q x^{2}-p y^{2})\dddot{h}(t)\right).
\end{eqnarray*}
They depend on four arbitrary functions $f, g, h$ and $k$ of variable $t$ in $C^\infty(\mathbb{R})$. Corresponding to the four arbitrary functions, we have operators
\begin{eqnarray}
&&Z_h=h(t)\partial_{t}+\frac{1}{2}\dot{h}(t)\left(x\partial_{x}+y \partial_{y}-u\partial_{u}-2v\partial_v\right)-\frac{qx^{2}-py^{2}}{16pq}\left(2 i  \ddot{h}(t)u\partial_{u}-\dddot{h}(t)\partial_{v}\right),\nonumber \\
&&X_f=f(t)\partial_{x}-\frac{1}{4p}x \left(2 u i\dot{f}(t)\partial_{u}-\ddot{f}(t)\partial_{v}\right),\nonumber\\
&& Y_g=g(t)\partial_{y}+\frac{1}{4q}y(2 u i\dot{g}(t)\partial_{u}-\ddot{g}(t)\partial_{v}), \nonumber \\
&&W_k=ik(t)u\partial_{u}-\frac{1}{2} \dot{k}(t)\partial_{v}.\label{e11}
\end{eqnarray}
These operators span an infinite dimensional Lie algebra, i.e., 2D-CNLS algebra, denoted it by $\mathfrak{L}=<X_f, Y_g, $ $Z_h, W_k>$, for arbitrary differential real functions $f=f(t), g=g(t), h=h(t)$ and $k=k(t)$ of class $C^\infty(\mathbb{R})$. Thus, the general element of the 2D-CNLS algebra $\mathfrak{L}$ can be written as
\begin{eqnarray}
V=X_f+Y_g+Z_h+W_k,\label{V}
\end{eqnarray}
and the commutation relations for $V\in \{X_F, Y_G, Z_H, W_K\}$ and $v\in \{X_f, Y_g, Z_h, W_k\}$ of 2D-CNLS algebra $\mathfrak{L}$ is given by
\begin{table}[h]
\caption{Commutator of 2D-CNLS algebra $\mathfrak{L}$}
\centering
\begin{tabular}{c| c c c c c}
\hline
$[V, v]$& $Z_h$& $X_f$& $Y_g$& $W_k$\\
\hline
$Z_H$ &$Z_{H\dot{h}-\dot{H}h}$ &$X_{H\dot{f}-\frac{1}{2}\dot{H}f} $ &$Y_{H\dot{g}-\frac{1}{2}\dot{H}g}$&$W_{H\dot{k}}$\\
$X_F$  &$X_{\frac{1}{2}F\dot{h}-\dot{F}h}$&$-\frac{1}{2p}W_{F\dot{f}-\dot{F}f}$ &0&$0$\\
$Y_G$  &$Y_{\frac{1}{2}G\dot{h}-\dot{G}h}$ &$0$ &$\frac{1}{2q}W_{G\dot{g}-\dot{G}g}$ &0\\
$W_K$  &$-W_{h\dot{K}}$ &$0$&$0$ &$0$\\
\hline
\end{tabular}
\end{table}

From the commutator, we notice that the 2D-CNLS algebra has a Levi decomposition
\begin{eqnarray}
\mathfrak{L}=\mathfrak{S}\oplus \mathfrak{N}, \label{Levi}
\end{eqnarray}
where $\mathfrak{N}=\{X_f, Y_g, W_k\}$ is a solvable ideal in $\mathfrak{L}$ and $\mathfrak{S}=\{Z_h\}$ is a simple Lie algebra which is isomorphic to the simple algebra \cite{Nath}
$${\mathfrak{J}}(\mathbb{R})=\{h(\xi)\partial_\xi|h(\xi)\in C^\infty(\mathbb{R})\},$$
of real vector fields on $\mathbb{R}$ defined by the mapping
$$\psi: {\mathfrak{J}}(\mathbb{R})\rightarrow S \textrm{ with } h(\xi)\partial_\xi\mapsto Z_{h(t)}.$$
Particularly, it is noticed that 2D-CNLS algebra $\mathfrak{L}$ can be written
\begin{eqnarray}
\mathfrak{L}=\mathfrak{L}_X+\mathfrak{L}_Y,\label{decomp}
\end{eqnarray}
as linear space, where $\mathfrak{L}_X=<Z_h, X_f, W_k>$ and  $\mathfrak{L}_Y=<Z_h, Y_g, W_k>$ are subalgebras of algebra $\mathfrak{L}$.

The following properties of the four operators in (\ref{e11}) are obvious and frequently used as a basic arithmetic operations in the computations of classifying the 2D-CNLS algebra.
\begin{lm}
Operators $X_f, Y_g, Z_h$ and $W_k$ are linear in their labeling functions $f, g, h$ and $k$ respectively and they are linear independent for not all zero functions $f, g, h$ and $k$.
\end{lm}

\subsection{The 2D-CNLS group}

It is well known that the symmetry groups of a differential equations are the solutions to the initial value problem of Lie equations of operator (\ref{e3}).

For our case, the 2D-CNLS groups are determined in the following two cases.

First, the function $h(t)\neq 0$ and $f(t), g(t)$ and $k(t)$ in (\ref{V}) are arbitrary. Solving the Lie equations
\begin{eqnarray}
&&\frac{dt^\prime}{d\varepsilon}={h(t^\prime)},\nonumber\\
&&\frac{dx^\prime}{d\varepsilon}={f(t^\prime) + \frac{1}{2} x^\prime \dot{h}(t^\prime)},\nonumber\\
&&\frac{dy^\prime}{d\varepsilon}={g(t^\prime) + \frac{1}{2} y^\prime \dot{h}(t^\prime)},\nonumber\\
&&\frac{du^\prime}{d\varepsilon }=\frac{1}{2}u^\prime\left(2ik(t^\prime)-\frac{i}{p}x^\prime  \dot{f}(t^\prime)+\frac{i}{q}y^\prime  \dot{g}(t^\prime)-\dot{h}(t^\prime)-\frac{i}{4pq}(q{x^\prime}^{2}-p{y^\prime}^2)\ddot{h}(t^\prime)\right), \nonumber \\
&&\frac{dv^\prime}{d\varepsilon}=-\frac{1}{2}\left(\dot{k}(t^\prime)-\frac{1}{2p}x^\prime \ddot{f}(t^\prime)+\frac{1}{2q} y^\prime \ddot{g}(t^\prime)+2v^\prime\dot{h}(t^\prime)-\frac{i}{8pq} (q{x^\prime}^{2}-p{y^\prime}^2)\dddot{h}(t^\prime)\right),\label{e18}
\end{eqnarray}
with initial values $t^\prime(0)=t,x^\prime(0)=x, y^\prime(0)=y, u^\prime(0)=u,v^\prime(0)=v$, we obtain the 2D-CNLS groups corresponding to 2D-CNLS algebra as the following
\begin{eqnarray}
&&t^\prime=\Phi^{-1}(\varepsilon +\Phi(t)),\no \\
&&x^\prime=[x+F(t^\prime,t)]\left[\frac{h(t^\prime)}{h(t)}\right]^{\frac{1}{2}},\no \\
&&y^\prime=[y+G(t^\prime,t)]\left[\frac{h(t^\prime)}{h(t)}\right]^{\frac{1}{2}},\no \\
&&u^\prime=u(t,x, y)\exp\left(\int^{t^\prime}\left(ik(t^\prime)-\frac{ i x^\prime \dot{f}(t^\prime)}{2p }+\frac{i y^\prime \dot{g}(t^\prime)}{2q}-\frac{ \dot{h}(t^\prime)}{2}-i\ddot{h}(t^{\prime})\left(\frac{{x^\prime}^{2} }{8p}-\frac{{y^\prime}^{2}}{8q}\right)\right)\frac{1}{h(t^\prime)} dt^\prime\right), \no \\
&&v^\prime=\left(\frac{h(t^\prime)}{h(t)}\right)^{-1}
\left\{v(t, x, y)-H(t^\prime,t) -\frac{1}{4h(t)}\left(\frac{x^\prime}{p}\left(\frac{\dot{h}(t^\prime)f(t^\prime)}{2h(t^\prime)}
-\dot{f}(t^\prime)\right)-\frac{y^\prime}{q}\left (\frac{\dot{h}(t^\prime)g(t^\prime)}{2h(t^\prime)}-\dot{g}(t^\prime)\right)\right)
 \right.\no \\
&&\hspace*{18mm}+\frac{1}{8h(t)}\left(\frac{h(t^\prime)^{-1}g(t^\prime)^{2}-h(t)^{-1}g(t )^{2}}{q}+\frac{h(t)^{-1}f(t)^{2}-h(t^\prime)^{-1}f(t^\prime)^{2}}{p}\right)\no \\
&&\hspace*{18mm}+\left.\frac{1}{16h(t)h(t^\prime)}\left(\frac{{x^\prime}^{2}}{p}
-\frac{{y^\prime}^{2}}{q}\right) \left(h(t^\prime)\ddot{h}(t^\prime)
-\frac{1}{2}\dot{h}(t^\prime)^{2}\right)\right\},\label{e19}
\end{eqnarray}
where
\begin{eqnarray}
&&\Phi(t)=\int_{t_{0}}^{t}\frac{1}{h(s)}ds,\,\,\, F(t^{'},t)=h(t)^{1/2}\int_{t}^{t^{'}}f(s)h(s)^{-3/2}ds,\no \\
&&G(t^{'},t)=h(t)^{1/2}\int_{t}^{t^{'}}g(s)h(s)^{-3/2}ds,\,\,\, H(t^{'},t)=h(t)^{-1}\int_{t}^{t^{'}}\frac{1}{2}\dot{k}(s)ds,\no
\end{eqnarray}
and $\Phi^{-1}$ denotes the inverse of $\Phi$ and $\varepsilon$ is the group parameter.

Next, we consider the case $h(t)\equiv 0$ and $f(t)$, $k(t)$ and $g(t)$ are arbitrary. Solving the corresponding initial value problem of the Lie equations, we obtain the 2D-CNLS groups as following
\begin{equation}
\left\{
\begin{array}{l}
t^{'}=t,\\
x^{'}=x+\varepsilon f(t),\\
y^{'}=y+\varepsilon g(t),\\
u^{'}=u\exp\left(\frac{i}{4pq}\varepsilon(4pq k(t)-q( 2x+\varepsilon f(t))\dot{f}(t)+p (2y+\varepsilon g(t))\dot{g}(t))\right),\\
v^{'}=v-\frac{\varepsilon}{8pq}\left(4pq \dot{k}(t)-q(2x+\varepsilon f(t))\ddot{f}(t)+p(2y+ \varepsilon g(t))\ddot{g}(t)\right).\label{e17}
\end{array}
\right.
\end{equation}

The immediate application of the symmetry groups (\ref{e19}) and (\ref{e17}) is to give us new solutions ($u^\prime, v^\prime$) from known ones ($u,v$). For instance, if we have a initial constant solution $u=a, v=b$, then (\ref{e19}) and (\ref{e17}) yield a family of solutions to 2D-CNLS system (\ref{e1}) depending on arbitrary functions $f(t), g(t), h(t), k(t)$ and constants $a$ and $b$.

It is obviously here that 2D-CNLS system (\ref{e1}) is also invariant under the reflections $R_x, R_y, R_{xy}$ and $R_u$ (discrete symmetries),
\begin{eqnarray}
&&R_x:t\mapsto t,x \mapsto -x, y \mapsto y, u \mapsto u, v \mapsto v,\nonumber\\
&&R_x:t\mapsto t,x \mapsto x, y \mapsto -y, u \mapsto u, v \mapsto v,\nonumber\\
&&R_{xy}:t\mapsto t,x \mapsto -x, y \mapsto -y, u \mapsto u, v \mapsto v,\nonumber\\
&&R_u:t\mapsto t,x \mapsto x, y \mapsto y, u \mapsto -u, v \mapsto v,\label{discrete}
\end{eqnarray}
which are not obtained from (\ref{V}) by solving its Lie equations.

\subsection{Some finite-dimensional subalgebras of physical transformations}
In the section, we show that the infinite dimensional 2D-CNLS algebra $\mathfrak{L}$ contains many specific subalgebras. They are interested in physical applications.

\textbf{1. First order subalgebras}

Taking the arbitrary functions $f(t), g(t), h(t)$ and $k(t)$ as the first-order polynomials in $t$, then we have
\begin{eqnarray*}
&&X_1=\partial_{x},\,\,\, Y_1=\partial_{y},\,\,\, Z_1=\partial_{t},\,\,\, W_1=iu\partial_{u},\\
&&X_t=t\partial_{x}-\frac{i}{2p}x u\partial_{u},\,\,\, Y_t=t\partial_{y}+ \frac{i}{2q}y u\partial_{u},\\
&&Z_t=\frac{1}{2}x\partial_{x}+\frac{1}{2}y\partial_{y} +t\partial_{t}-\frac{1}{2}u\partial_{u}-v\partial_{v},\,\,\, W_t=itu\partial_{u}-\frac{1}{2}\partial_{v},
\end{eqnarray*}
as a basis of an eight-dimensional  solvable Lie algebra $\mathfrak{L}_8=<X_t, Y_t, Z_t, W_t, X_1, Y_1, Z_1,$ $ W_1>$. It has a seven-dimensional nilpotent ideal $\mathfrak{N}=<X_t, Y_t, W_t, X_1, Y_1, Z_1, W_1>$.

\textbf{2. Quadratic subalgebras}

\textbf{a. }A three-dimensional subalgebra is obtained from $Z(h)$ by restricting $h=h(t)$ into quadratic polynomials in $t$. This yields
\begin{eqnarray}
&&Z_1=\partial_t,\,\,\, Z_t=\frac{1}{2}x\partial_{x}+\frac{1}{2}y\partial_{y}+t\partial_{t} -\frac{1}{2}u\partial_{u}-v\partial_{v},\nonumber\\
&&Z_{t^2}=xt\partial_{x}+yt\partial_{y}+t^2\partial_{t} -(t+\frac{qx^2-py^2}{4pq}i)u\partial_{u}-2tv\partial_{v}.\label{sp}
\end{eqnarray}
Their commutation relations are $[Z_1,Z_{t^2}]=2Z_t, [Z_t,Z_{t^2}]=Z_{t^2}, [Z_1,Z_t]=Z_1$, so we obtain the algebra $\mathfrak{L}_s=<Z_1,Z_t,Z_{t^2}>$ which is isomorphism to sl$(2,\mathbb{R})$.  The $Z_{t^2}$ generates the projective symmetry group
\begin{eqnarray}
&&x^{'}=x/(1-\varepsilon t), \,\, y^{'}=y/(1-\varepsilon t), \,\,t^{'}=t/(1-\varepsilon t),\nonumber\\ &&u^{'}=(1-\varepsilon t)u\exp({i(py^{2}-qx^{2})\varepsilon}/{4pq(1-\varepsilon t)}),\,\,
v^{'}=(1-\varepsilon t)^{2}v.\nonumber
\end{eqnarray}

\textbf{b. }Another three-dimensional algebra is obtained by restricting $k=k(t)$ in $W_k$ to quadratic polynomials in $t$ and obtain the basis $W_1=iu\partial_u, W_t=itu\partial_u -\frac{1}{2}\partial_{v}, W_{t^2}=it^2u\partial_{u}-t\partial_{v}$ with $[W_1, W_t]= [W_1, W_{t^2}]=[W_t, W_{t^2}]=0$. This is a three-dimensional Abelian Lie algebra. In fact, for any an integer $n$, we can obtain $n$-dimensional Abelian algebra from $W_k$ by taking the $k=k(t)$ as an $n-1$-degree polynomials in $t$ since the Abelian property of operator $W_k$ for arbitrary functions $k$.

\textbf{3. An affine loop algebra}

One interesting aspect of the 2D-CNLS algebra is that the two subalgebras $\mathfrak{L}_X$ and $\mathfrak{L}_Y$ in (\ref{decomp}) can be simultaneously embedded into an affine loop algebra if the arbitrary functions contained in the algebras are taken as Laurent polynomials. Indeed, let us consider the subalgebra $\mathfrak{L}^n$ of the 2D-CNLS algebra obtained by restricting the functions $f, g, h$ and $k$ in (\ref{e11}) to be Laurent polynomials in $t$. A basis for this subalgebra is given by
\begin{eqnarray}
&&X_{t^n}=t^n\partial_x-\frac{i}{2p}nt^{n-1}x u\partial_u+\frac{1}{4p}n(n-1)t^{n-2}x\partial_v,\no\\
&&Y_{t^n}=t^n\partial_y+\frac{i}{2q}nt^{n-1}y u\partial_u-\frac{1}{4q}n(n-1) t^{n-2}y\partial_v,\no\\
&&Z_{t^n}=t^{n}\partial_t+\frac{1}{2}nt^{n-1}(x\partial_x+y\partial_y
-u\partial_u-2v\partial_v)-\frac{qx^2-py^2}{16pq}n(n-1)\left(2uit^{n-2}\partial_u
-(n-2)t^{n-3}\partial_v\right),\no\\
&&W_{t^n}=it^nu\partial_u-\frac{1}{2}nt^{n-1}\partial_v,\nonumber
\end{eqnarray}
where $n\in \mathbb{N}$ (Integer set).

The Table 2 in the following is the commutation relations between $\{X_{t^n}, Y_{t^n}, Z_{t^n}, W_{t^n}\}$ and $\{X_{t^m}, Y_{t^m},$ $Z_{t^m}, W_{t^m}\}$ derived from Table 1.
\begin{table}[hhh]
\caption{Commutators of $\{X_{t^i},Y_{t^i},Z_{t^i}, W_{t^i}\}$ for $i=n, m$}
\centering
\begin{tabular}{c| c c c c c}
\hline
$[\cdot, \cdot]$& $Z_{t^m}$& $X_{t^m}$& $Y_{t^m}$& $W_{t^m}$\\
\hline
$Z_{t^n}$ &$(m-n)Z_{t^{m+n-1}}$ &$(m-\frac{1}{2}n)X_{t^{m+n-1}}$ &$(m-\frac{1}{2}n)Y_{t^{m+n-1}}$ &$mW_{t^{n+m-1}}$\\
$X_{t^n}$  &$(\frac{1}{2}m-n)X_{t^{m+n-1}}$&$-\frac{m-n}{2p}W_{t^{n+m-1}}$ &0&$0$\\
$Y_{t^n}$  &$(\frac{1}{2}m-n)Y_{t^{m+n-1}}$ &$0$ &$\frac{m-n}{2q}W_{t^{n+m-1}}$ &0\\
$W_{t^n}$  &$-nW_{t^{n+m-1}}$ &$0$&$0$ &$0$\\
\hline
\end{tabular}
\end{table}

Let $\mathfrak{L}_X^n=<X_{t^n}, Z_{t^n}, W_{t^n}>$ and $\mathfrak{L}_Y^n=<Y_{t^n}, Z_{t^n}, W_{t^n}>$, then $\mathcal{L}^n=\mathfrak{L}_X^n+\mathfrak{L}_Y^n$ as linear space.

Let us now consider the eleven-dimensional Lie algebra $\mathfrak{L}_{11}$ generated by the following vector fields:
\begin{eqnarray*}
&&\Delta=(1/2)(x\partial_x+y\partial_y-u\partial_u-2v\partial_v);\,\,\, U=i u\partial_u;\,\,\, V=-(1/2)\partial_v;\\
&&A=((py^2-qx^2)/8pq)U;\,\,\,P=((py^2 -qx^2)/8pq)V;\,\,\,U_x=(-x/2p)U;\\
&&V_x=(-x/2p)V;\,\,\,U_y=(y/2q)U;\,\,\,V_y=(y/2q)V, \,\,\,X=\partial_x;
\,\,\,Y=\partial_y.
\end{eqnarray*}
This is a solvable algebra and its nilradical is spanned by
$\{A, P, X, Y, U, V, U_x, U_y, V_x, V_y\}$ and it contains an eight dimensional Abelian ideal spanned by $\{A, P, U, V, U_x, U_y, V_x, V_y\}$. Obviously, algebra $\mathfrak{L}_{11}$ is not a subalgebra of 2D-CNLS algebra. We can write
\begin{eqnarray}
\mathfrak{L}_{11}=\mathfrak{L}_x+\mathfrak{L}_y,
\end{eqnarray}
as a linear space, where $\mathfrak{L}_x=<\Delta, X, U, V, A, P, U_x, V_x>$ and $\mathfrak{L}_y=<\Delta, Y, U, V, A, P, U_y, V_y>$. Both $\mathfrak{L}_x$ and $\mathfrak{L}_y$ is solvable subalgebras containing five- dimensional Abelian ideals spanned by $<V, A, P, U_x, V_x>$ and $<V, A, P, U_y, V_y>$ respectively, with common center $<U>$. The two subalgebras should be embedded into simple Lie algebras. TIt is known that  algebra $A_4$ or in our case sl(5,$\mathbb{R}$) is a simple Lie algebra of lowest dimension that contains a five dimensional Abelian subalgebra. Indeed, the following traceless matrices with eight parameters $\delta, \rho, \gamma, \nu, \alpha, \mu, \sigma$ and $\kappa$
\begin{equation}       
\mathcal{A}=\left(
\begin{array}{ccccc}
 \frac{11 }{10}\delta & 0 & -\rho & 2 \nu & \upsilon \\
 0 & \frac{1}{10}\delta & -\alpha & 2 \mu & \sigma \\
 0 & 0 & -\frac{9 }{10}\delta & \kappa & 0 \\
 0 & 0 & 0 & -\frac{2 }{5}\delta & \frac{1}{4 p}\kappa \\
 0 & 0 & 0 & 0 & \frac{1}{10} \delta \\
\end{array}
\right)            
\end{equation}
provide a representation of the Lie algebra $\mathfrak{L}_x$. Each entry in the basis of $\mathfrak{L}_x$ corresponds to a matrix obtained from $\mathcal{A}$ in which one parameter is taken 1 and all other are zero. For instance, the $\Delta$ corresponds to the matrix by setting $\delta=1$ and all other parameters equal to zero in $\mathcal{A}$. Similarly for $X, U, V, A, P, U_x, V_x$, we have correspondences: $X\rightarrow\kappa, U\rightarrow\sigma, V\rightarrow\upsilon, A\rightarrow\alpha, P\rightarrow\rho, U_x\rightarrow\mu, V_x\rightarrow\nu$. By the same manner, for $\mathfrak{L}_y$, we have the same form matrix representatives obtained by replacing the $p$ in $\mathcal{A}$ by the $q$.

Let us now define a natural grading on $\mathfrak{L}_x$ ($\mathfrak{L}_y$) by attributing the degree ``$\mu$" to entries of the basis of $\mathfrak{L}_x$ ($\mathfrak{L}_y$). The degree is equal to the distance between, in the same column, the corresponding parameter term and the diagonal term of $\mathcal{A}$. Thus $\Delta$ has degree 0, $A$ and $X$ ($Y$) degree 1, $P$ and $U_x$ ($U_y$) degree 2, $U$ and $V_x (V_y)$ degree  3, $V$ degree 4. Consequently, $\mathfrak{L}_{11}=\mathfrak{L}_x +\mathfrak{L}_y$ embedded in to sl$(5,\mathbb{R})$. Meantime, each element in the basis of $\mathfrak{L}_X^n$ ($\mathfrak{L}_Y^n$) has a well-defined degrees (sum of degrees in $t$ and entry of $X, U, V, A, P, U_x$ ($U_y$) and $V_x$($V_y$) just defined above), namely $n-1, n+1$, and $n+3$.

From the embedding constructed above for $\mathfrak{L}_{11}$ through $\mathfrak{L}_x$ and $\mathfrak{L}_y$ into sl$(5, \mathbb{R})$ and from the representations of $\mathfrak{L}^n$ given below
\begin{eqnarray}
&&Z_{t^n}=nt^{n-1}\Delta+n(n-1)t^{n-2}A+n(n-1)(n-2)t^{n-3}P+t^n\partial_t,\nonumber\\
&&Y_{t^n}=t^nY+nt^{n-1}U_y+n(n-1)t^{n-2}V_y,\nonumber\\
&&X_{t^n}=t^nX+nt^{n-1}U_x+n(n-1)t^{n-2}V_x,\nonumber\\
&&W_{t^n}=t^nU+nt^{n-1}V,\nonumber
\end{eqnarray}
we see that $\mathfrak{L}^n$ is a subalgebra of the affine loop algebra $A_{4}^{(1)}$ defined by
\begin{eqnarray*}
A_{4}^{(1)}:=\{R[t,t^{-1}]\otimes \textrm{sl}(5,R)\}\oplus R[t,t^{-1}]\frac{d}{dt}.
\end{eqnarray*}

The Levi decomposition (\ref{Levi}) also holds for $\mathfrak{L}^n$. Indeed, from the commutation relation in Table 1 we see that $\mathfrak{N}=\{X_{t^n}, Y_{t^n}, W_{t^n}\}$ forms a nilpotent ideal. The element $Z_{t^n}$ form a Lie algebra $\mathfrak{S}$ isomorphic to the simple $\mathbb{N}$-graded algebra $\mathfrak{L}_t=\mathbb{R}[t, t^{-1}]\frac{d}{dt}$. A basis for $\mathfrak{L}_t$ is given by the set of derivations $(d_n)_{n\in\mathbb{N}}$ given by $d_n=t^n\frac{d}{dt}$ with commutation relations
$$[d_n, d_m]=(m-n)d_{m+n-1}.$$

From the above discussion, we see that the 2D-CNLS system (\ref{e1}) admits rich finite and infinite algebras as well as physical symmetries.  In section four, we will give lower dimensional subalgebra classification under the adjoint action of the 2D-CNLS group, which leads to a list of distinguishable equivalent classess of finite dimensional subalgebras of the 2D-CNLS algebra. Consequently, it will clarify the how many equivalent classes of invariant solutions exist for the 2D-CNLS system (\ref{e1}).

\section{Adjoint action of 2D-CNLS group}
\subsection{Adjoint action of 2D-CNLS group}
In convenience, let $z=(t, x, y, u, v)$ and $\partial_z=(\partial_t, \partial_x, \partial_y, \partial_u, \partial_v)$.

Let $\Gamma: z\rightarrow \hat{z}$ be a symmetry of system (\ref{e1}) and
$\mathcal{X}=z\cdot\partial_z\in \mathfrak{L}$ be the generator of an one-parameter symmetry group of system (\ref{e1}) mapping $z$ to $z^\prime$. The dot represents the scalar production of $z$ and $\partial_z$ as being seen vectors. Thus, under transformation $\Gamma$, we conclude that the  transformed operator $\hat{\mathcal{X}}$ of $\mathcal{X}$ by
\begin{eqnarray}\hat{\mathcal{X}}=\Gamma \mathcal{X} \Gamma^{-1}=\hat{z}^\prime\cdot\partial_{\hat{z}}, \label{Y}\end{eqnarray} is the generator for the conjugate group of the group generated by $\mathcal{X}$ with respect to $\Gamma$ with $\hat{z}^\prime=\mathcal{X}(\hat{z})=(\mathcal{X}(\hat{t}), \mathcal{X}(\hat{x}), \mathcal{X}(\hat{y})$, $\mathcal{X}(\hat{u}), X(\hat{v}))$.

In particular, if the symmetry $\Gamma$ is taken as an element $\Gamma_\delta$ of 2D-CNLS group with a parameter $\delta$ generated by $\mathcal{Z}\in \mathfrak{L}$, then (\ref{Y}) induces the inner automorphism group $\exp(\delta\cdot ad(\mathcal{Z}))$ of 2D-CNLS algebra $\mathfrak{L}$ (adjoint representation of 2D-CNLS group). The action induced by the inner automorphism group on the Lie algebra $\mathfrak{L}$ is called the adjoint action of the 2D-CNLS group. The conjugating between $\mathcal{X}$ and $\hat{\mathcal{X}}$ is implemented by formula given in \cite{B2} as
\begin{eqnarray}\hat{\mathcal{X}}=\exp(\delta\cdot ad(\mathcal{Z}))\mathcal{X}=\sum_{k=0}^\infty\frac{\delta^k}{k!}ad(\mathcal{Z})^k \mathcal{X},\label{YG2}\end{eqnarray}
where $ad(\mathcal{Z})$ is the adjoint operator of $\mathcal{Z}$ defined by $ad(\mathcal{Z})\mathcal{Y}=[\mathcal{Y},\mathcal{Z}]$ with Lie product $[\cdot, \cdot]$ of $\mathfrak{L}$ for all $\mathcal{Y}\in \mathfrak{L}$. We call that two operators $\mathcal{X}$ and $\hat{\mathcal{X}}$ are  equivalent if (\ref{YG2}) is satisfied for some $\mathcal{Z}\in \mathfrak{L}$ and denote it as $\mathcal{X} \sim \hat{\mathcal{X}}$. This is an equivalent relationship on the algebra $\mathfrak{L}$. Hence it yields equivalent classification of 2D-CNLS algebra $\mathfrak{L}$.

\subsection{Properties of adjoint action of 2D-CNLS group}
Based on the commutator Table 1 and formulae (\ref{Y}) and (\ref{YG2}), the adjoint actions $\exp(\delta\cdot ad(V))v$ of 2D-CNLS group with $V\in \{X_F, Y_G, Z_H, W_K\}$ on $v\in\{X_f, Y_g, Z_h, W_k\}$ are given in Table 3 in which $\widetilde{f}=f(t)\left(\frac{H(t^\prime)}{H(t)}\right)^{\frac{1}{2}}$,  $\widetilde{g}=g(t)\left(\frac{H(t^\prime)}{H(t)}\right)^{\frac{1}{2}}$,  $\widetilde{h}=h(t)\frac{H(t^\prime)}{H(t)}$, $\overline{F}=\frac{1}{2p}(F\dot{f}-\dot{F}f)$, $\overline{G}=\frac{1}{2q}(G\dot{g}-\dot{G}g)$, $\widehat{F}=\frac{1}{4p}(\frac{1}{2}F\dot{h}-\dot{F}h)$, $\widehat{G}=\frac{1}{4q}(\frac{1}{2}G\dot{h}-\dot{G}h)$ and  $\widetilde{k}=\widetilde{k}(t^\prime)=k(t)$ with $t=\Phi(-\delta+\Phi(t^\prime))$.
\begin{table}[hhh]\label{ad}
\caption{Adjoint actions of the 2D-CNLS group on $\mathfrak{L}$}
\centering
\begin{tabular}{c| c c c c c}
\hline
$\exp(\delta\cdot ad(V))v$& $Z_h$& $X_f$& $Y_g$& $W_k$\\
\hline
$\exp(\delta\cdot ad(Z_H))$ &$Z_{\widetilde{h}}$ &$X_{\widetilde{f}} $ &$Y_{\widetilde{g}}$&$W_{\widetilde{k}}$\\
$\exp(\delta\cdot ad(X_F))$  &$Z_h-4p\delta X_{\widehat{F}} -\delta^2 W_{F\dot{\widehat{F}}-\dot{F}\widehat{F}}$&$X_f+\delta W_{\overline{F}}$&$Y_g$ &$W_k$\\
$\exp(\delta\cdot ad(Y_G))$  &$Z_h-4q\delta Y_{\widehat{G}}+\delta^2 W_{G\dot{\widehat{G}}-\dot{G}\widehat{G}}$ &$X_f$ &$Y_g-\delta W_{\overline{G}}$ &$W_k$\\
$\exp(\delta\cdot ad(W_K))$  &$Z_h+\delta W_{h\dot{K}}$ &$X_f$&$Y_g$ &$W_k$\\
\hline
\end{tabular}
\end{table}

We frequently use the following three Theorems as basic operations to normalize the elements of Lie algebra $\mathfrak{L}$.
\begin{thh}\label{thh1} The symmetry group generated by generator $Z_H$ in {\rm 2D-CNLS} algebra $\mathfrak{L}$ for a differential nonzero function $H=H(t)$ is given by
\begin{eqnarray*}
&&t^\prime=\phi^{-1}(\delta+\phi(t)),\\
&&x^\prime=x(\frac{H(t^\prime)}{H(t)})^{\frac{1}{2}},\\
&&y^\prime=y(\frac{H(t^\prime)}{H(t)})^{\frac{1}{2}},\\
&&u^\prime=u(t,x,y)\exp\left(-\frac{1}{2}\int^{t^\prime}\left( \frac{ \dot{H}(t^\prime)+\frac{i}{4pq}(q{x^\prime}^2
-p{y^\prime}^2)\ddot{H}(t^\prime)}{H(t^\prime)}\right)dt^\prime\right),\\
&&v^\prime=\left(\frac{H(t^\prime)}{H(t)}\right)^{-1}
\left\{v(t, x, y) +\frac{1}{16pqH(t)H(t^\prime)}\left(q{x^\prime}^{2}-p{y^\prime}^{2}\right) \left(H(t^\prime)\ddot{H}(t^\prime)-\frac{1}{2}\dot{H}(t^\prime)^{2}\right)
\right\},
\end{eqnarray*}
where $\Phi(t)=\int^{t}\frac{1}{H(s)}ds$ and $\delta$ is group parameter.
\end{thh}

Proof: This is obtained from (\ref{e19}) by setting $h=H$ and $f=g=k=0$.

\begin{thh}\label{thh2}If $h\not =0$ in $V=Z_h+X_f+Y_g+W_k\in \mathfrak{L}$, then $V$ is equivalent to $Z_1$ for arbitrary functions $h, f, g$ and $k$ of variable $t$, {\rm i.e.}, $V\sim Z_1$.
\end{thh}

Proof: Successively using adjoint actions $\exp(\delta\cdot ad (X_F)), \exp(\lambda\cdot ad(Y_G))$ and $\exp(\nu\cdot ad (W_K))$ on $V$ and choosing suitable $F, G, K, \delta, \lambda, \nu$ for given $f, g, k$, we can normalize $V$ into $Z_1$ by the following steps.

\textbf{Step 1}. According to Table 3, we first can choose $G$ so as to transform $g$ away from $V$ by acting on it with $\exp(\delta\cdot ad (Y_G))$. Indeed, it can be verified that if we take $G$ satisfying $g-\delta(G\dot{h}/2-\dot{G}h)=0$, i.e.,
\begin{eqnarray}
G(t)=\sqrt{h(t)} \left(a \int^tg(s)h(s)^{-3/2}ds+c\right),\nonumber
\end{eqnarray}
where $a\not=0$ and $c$ are arbitrary constants, as the function labeling the element $Y_G$ of the 2D-CNLS algebra and $\delta=-a^{-1}$ as the value of group parameter $\delta$, then we have $\exp(\delta\cdot ad(Y_G))V=Z_h+X_f+W_k$, where actually $k$ is a new function depends on the chosen $G$ and we denote it again $k$.

\textbf{Step 2.} The next step is to choose the function $F$ so as to transform $f$ away from $Z_h+X_f+W_k$ by performing on it with $\exp(\delta\cdot ad(X_F))$. Again if we take $F$ satisfying $f-\delta(F\dot{h}/2-\dot{F}h)=0$, i.e.,
\begin{eqnarray}
F(t)=\sqrt{h(t)} \left(a \int^tf(s)h(s)^{-3/2}ds+c\right),\nonumber
\end{eqnarray}
where $a\not=0$ and $c$ are arbitrary constants, as the function labeling the element $X_F$ of the 2D-CNLS algebra and $\delta=-a^{-1}$ as the value of group parameter $\delta$, then we have $\exp(\delta\cdot ad(X_F))(Z_h+X_f+W_k)=Z_h+W_k$.

\textbf{Step 3. }The third step is to choose the function $K$ so as to drop $k$ away from $Z_h+W_k$ by acting on the $Z_h+W_k$ with $\exp(\delta\cdot ad(W_K))$. Again if we take $K$ satisfying $k+\delta h \dot{K}=0$, i.e.,  $K(t)=a\int^tk(s)h(s)^{-1}ds+c$ with $a=-\delta^{-1}$, then we have
\begin{eqnarray*}
\exp(\delta\cdot ad(W_K))(Z_h+W_k)=Z_h.
\end{eqnarray*}

\textbf{Step 4. }The last step is to choose the function $H$ so as to normalize $h$ to 1 in $Z_h$ by acting on it with adjoint transformation $\exp(\delta\cdot ad(Z_H))$. Indeed,
$\exp(\delta \cdot ad(Z_H))Z_h=Z_{\widetilde{h}}$ by formulas (\ref{Y}) and  $\widetilde{h}=h(t)\frac{H(t^\prime)}{H(t)}$ with $t=\Phi^{-1}(-\delta+\Phi(t^\prime))$.
It is shown in \cite{main1} and references there that there exists a function $H$ satisfying the relation
\begin{eqnarray}
h(t)\frac{H(t^\prime)}{H(t)}=1.\label{para1}
\end{eqnarray}

Let $H$ be a solution to (\ref{para1}). Then we normalize $h$ to 1. Dropping primes, we have
\begin{eqnarray*}\exp(\delta \cdot ad(Z_H))Z_h=Z_1.\end{eqnarray*}

This has proved $V \sim Z_1$.

\begin{thh}\label{thh3} For arbitrary functions $f, g$ and $k$ of $t$, we have

{\rm a}. if $f\not=0$, then $X_f+Y_g+W_k\sim X_1+Y_{\widetilde{g}}$ for an arbitrary function $\widetilde{g}$ of $t$;

{\rm b}. if $g\not=0$, then $X_f+Y_g+W_k\sim Y_1+X_{\widetilde{f}}$ for an arbitrary function $\widetilde{f}$ of $t$;

{\rm c}. if $fg\not=0$, then there exist functions $\widetilde{f}, \widetilde{g}$ of $t$ such that $X_f+Y_g+W_k\sim X_{\widetilde{f}}+Y_1\sim X_1+Y_{\widetilde{g}}$.
\end{thh}

Proof: The proof of a) and b) is included in the proof of c). Hence we give the proof of c) in the following.

As suggested by the commutation relations in Table 1 for the 2D-CNLS algebra, we can first choose $F$ so as to eliminate $k$ from $V$ by acting on it with $\exp(\delta\cdot ad(X_F))$. Since
$$\exp(\delta\cdot ad(X_F))V=X_f+Y_g+W_{k+\frac{\delta}{2p}(F\dot{f}-\dot{F}f)}.$$
taking $k+\frac{\delta}{2p}(F\dot{f}-\dot{F}f)=0,$ i.e.,
$$F(t)=f(t) \left(2a p \int^t k(s) f(s)^{-2}ds+c\right),$$
with $a=\delta^{-1}$ and arbitrary constant $c$, we transform $V$ into $V^\prime=X_f+Y_g$. Then, applying the adjoint map $\exp(\delta\cdot ad(Z_H)))$ and the symmetry generated by $Z_H$ (see Theorem \ref{thh1}) on $V^\prime$, we have
$$\exp(\delta\cdot ad(Z_H)))V^\prime=X_{\widetilde{f}}+Y_{\widetilde{g}},$$
where $\widetilde{f}=f(t)\left(\frac{H(t^\prime)}{H(t)}\right)^{\frac{1}{2}}, \,\, \widetilde{g}=g(t)\left(\frac{H(t^\prime)}{H(t)}\right)^{\frac{1}{2}}$ with $t=\Phi^{-1}(-\delta+\Phi(t^\prime))$.

Like discussion about the existence of function $H$ in (\ref{para1}), we can set either $\widetilde{f}=1$  or $\widetilde{g}=1$ for a selection of $H$. Consequently, we have the conclusions of the theorem.

\subsection{General structures of two-, three- dimensional Lie subalgebras}

The following two lemmas are theoretical bases for algebra classification.
\begin{lm}\label{lm1}
Any two- dimensional Lie algebra with a basis $\{A_1, A_2\}$  over $\mathbb{R}$ is isomorphic to one of the following ones:

{\rm (1)}Abelian: $[A_1, A_2]=0.$

{\rm (2)}Solvable, non- Abelian: $[A_1, A_2]=A_1$.
\end{lm}

\begin{lm}\label{lm2}{\rm \cite{main1, main2} }
Any three- dimensional Lie algebra with a basis $\{A_1, A_2, A_3\}$  over  $\mathbb{R}$ can be either simple or solvable. In simple case, the algebra is isomorphic to \emph{sl(2, } $\mathbb{R}$\emph{)} with a two dimensional non Abelian suabalgebra $<A_1,A_2>$ with the commutations
\begin{eqnarray}
[A_1, A_2]=A_1, \,\,[A_1, A_3]=2A_2, \,\,[A_2, A_3]=A_3.\label{sl}
\end{eqnarray}
In the solvable case, it always will have a two- dimensional Abelian ideal, i.e., there a basis $\{A_1, A_2, A_3\}$ satisfying
\begin{equation}
[A_1, A_2]=0,\,\, [A_1,A_3]=a A_1+bA_2,\,\, [A_2,A_3]=cA_1+dA_2,\label{three}
\end{equation}
\end{lm}

In the isomorphism classification of the three dimensional Lie algebra over $\mathbb{R}$, the matrix
M={\footnotesize$\left[
  \begin{array}{cc}
    a & b \\
    c & d \\
  \end{array}\right]$
}from (\ref{three}) have six standard forms, i.e.,
{\footnotesize\begin{eqnarray}
 M_1=\left[
  \begin{array}{cc}
    0 & 0 \\
    0 & 0 \\
  \end{array}\right], M_2=\left[
  \begin{array}{cc}
    1& 0 \\
    0& 0 \\
  \end{array}\right], M_3=\left[
  \begin{array}{cc}
    0 & 0 \\
    1 & 0 \\
  \end{array}\right], M_4=\left[
  \begin{array}{cc}
    1 & 0 \\
    0 & \alpha \\
  \end{array}\right], M_5=\left[
  \begin{array}{cc}
    \alpha & 1 \\
    -1 & \alpha \\
  \end{array}\right], M_6=\left[
  \begin{array}{cc}
   1 &0 \\
   1 &1 \\
  \end{array}\right].\label{stand}
\end{eqnarray}
}

The details of the classification of three dimensional Lie algebra are seen in \cite{thes}.
\section{Classifications of Low Dimensional Lie Subalgebras}
In this section we give the classification of all  one-, two-, and three- dimensional subalgebras of the 2D-CNLS algebra under the adjoint action of the 2D-CNLS group.
\subsection{One-dimensional subalgebras}
We will show that there are four conjugacy classes of one-dimensional subalgebras of the 2D-CNLS algebra under the adjoint action of the 2D-CNLS group, with representatives $Z_1, X_1, X_\phi+Y_1$ and $W_\psi$, respectively, for arbitrary differential functions $\phi$ and $\psi$ of $t$ in $C^\infty(\mathbb{R})$. The approach we take is similar to the one used in \cite{main1, main2}. The focus here is that one obtains differential conditions on the arbitrary functions labeling the group elements, as seen in proof of Theorem \ref{thh2}. The adjoint action is used to turn the generators of the subalgebras into normal forms.

There are three cases to be considered in the classification of the one- dimensional subalgebras of the 2D-CNLS algebra, generated by typical elements of the form (\ref{V}) into conjugacy classes under the adjoint group of the 2D-CNLS group.

\textbf{Case A: $h\not=0$.} By Theorems \ref{thh1}-\ref{thh2}, we claim that general element $V=Z_h+X_f+Y_g+W_k$ always can be normalized to $Z_1$ by  the adjoint group of the 2D-CNLS group. Hence, we have a subalgebra $\mathfrak{L}_{1,1}=<Z_1>$.

\textbf{Case B: }$h=0$ and $f^2+g^2\not=0$. Using Theorem \ref{thh3}, we claim that $V=X_f+Y_g+W_k$ can be transformed into $X_1$ or $X_f+Y_1$ or $Y_1$ by the adjoint group of the 2D-CNLS group. That is,

For the case $f\not=0, g=0$, we have an essential subalgebra $\mathfrak{L}_{1,2}=<X_1>$.

If $fg\not=0$, we have subalgebra $\mathfrak{L}_{1,3}=<X_\phi+Y_1>$ for an arbitrary function $\phi\not=0$.

For the case $f=0$ and $g\not=0$, we can prove $V \sim Y_1$. This can be merged into the $\mathfrak{L}_{1,3}$ by allowing $\phi=0$.

\textbf{Case C:} $h=f=g=0$ and $k\not=0.$  Because of the invariance of $W_k$ under adjoint actions of 2D-CNLS group, we have subalgebra $\mathfrak{L}_{1, 4}=<W_\psi>$ for the arbitrary nonzero function $\psi$.

To summarize, we have proved the following theorem.
\begin{thh}\label{thh4} an arbitrary one - dimensional subalgebra of {\rm 2D-CNLS} algebra $\mathfrak{L}$ is conjugate, under adjoint group of {\rm 2D-CNLS} group, to precisely one of the following.
\begin{eqnarray}\mathfrak{L}_{1,1}=<Z_1>,\,\, \mathfrak{L}_{1,2}=<X_1>,\,\, \mathfrak{L}_{1,3}=<X_\phi+Y_1>,\,\, \mathfrak{L}_{1,4}=<W_\psi>,\label{o-d}\end{eqnarray}
for arbitrary differential functions $\phi$ and $\psi\not=0$.
\end{thh}

\subsection{Two-dimensional subalgebras}
By Lemma 2, there exist two types of two- dimensional Lie algebras both over $\mathbb{R}$, namely Abelian and solvable non-Abelian algebras spanned by a basis $\{A_1, A_2\}$, satisfying $[A_1, A_2]=A_1$.

We will take $A_1$ in one of the four possible forms obtained in Theorem \ref{thh4}, and let $A_2=Z_h+X_f+Y_g+W_k$ be a general element of the 2D-CNLS algebra. We first impose the commutation relations, then simplify $A_2$ using the stabilizer (isotropy) group of $A_1$ in the adjoint group of 2D-CNLS algebra.

\subsubsection{Abelian algebras}
(1.1)$A_1=Z_1$.
Requiring $[A_1, A_2]=0$ and using Table 1 we find $\dot{h}=\dot{f}=\dot{g}=\dot{k}=0$. Hence $A_2=aZ_1+bX_1+cY_1+dW_1$. Replacing $A_2$ by $A^\prime_2=A_2-aA_1$ we effectively set $a=0$ in $A_2$. Subsequently using $\exp(\lambda\cdot ad(W_t))$, $\exp(\delta\cdot ad (X_t))$  and $\exp(\sigma\cdot ad (Y_t))$ on $A_1=Z_1$ and $A_2=bX_1+cY_1+dW_1$, we normalize $A_2$ and keep the invariance of $A_1$.

After applying these groups, we have
\begin{eqnarray}
&&A_1\sim Z_1+\delta X_1+\sigma Y_1+(\lambda-\frac{1}{4q}(\delta^2\epsilon-\sigma^2))W_1,\nonumber \\
&&A_2\sim b X_1+c Y_1+(d-\frac{1}{2q}(b\delta\epsilon-c\sigma))W_1.\label{tw1}
\end{eqnarray}
Let
\begin{eqnarray}\lambda-\frac{1}{4q}(\delta^2\epsilon-\sigma^2))=0,\,\,\, d-\frac{1}{2q}(b\delta\epsilon-c\sigma))=0, \,\,\,\delta=A b, \,\,\, \sigma= A c, \label{tw2}
\end{eqnarray}
for a nonzero constant $A$.

Suppose $d\not=0$, then it implies $c^2\not=\epsilon b^2$. Hence system (\ref{tw2}) has solution $\lambda=\frac{1}{2} d A , \delta=A b, \sigma= A c$ with $A=2dq/(\epsilon b^2-c^2)$. Further we suppose $b\not=0$, then we get equivalent algebras $<A_1, A_2>$ $\sim <Z_1+A(bX_1+cY_1), bX_1+cY_1>$ from (\ref{tw1}). By scaling, we can take $A_1=Z_1$ and $A_2=X_1+\alpha Y_1$. Therefore, we have $\mathfrak{L}_{2,1}=<Z_1, X_1+\alpha Y_1>$ for arbitrary constant $\alpha$. Similarly, if $b=0, c\not=0$, then we get $\mathfrak{L}_{2,2}=<Z_1, Y_1>$. If $b=c=0$, we get $\mathfrak{L}_{2,3}=<Z_1, W_1>$.

In the case where $d=0$, the two- dimensional subalgebra belongs to $\mathfrak{L}_{2,1}$ or $\mathfrak{L}_{2,2}$.

(1.2)$A_1=X_1$.

We again take $A_2=Z_h+X_f+Y_g+W_k.$ The condition $[A_1, A_2]=0$ implies  $\dot{f}=\dot{h}=0$ and $A_2=aZ_1+bX_1+Y_g+W_k$. We eliminate $b$ by linear combination with $A_1$. If $a\not=0$, conjugating by $\exp(\lambda ad(Y_G)+\nu ad (W_k))$ and appropriately choosing the $F, G, K, \delta, \nu$ and $\lambda$ like doing in Theorem \ref{thh2}, we can arrive at $A_2=Z_1$. This can be merged into  the subalgebra $\mathfrak{L}_{2,1}$ by setting $\alpha=0$. If $a=0$ and $g\not=0$, conjugating $A_2$ by $\exp(\delta\cdot ad(Y_G))$ and selecting $G$ satisfying $(G\dot{g}-\dot{G}g)/(2q)=k$, we can transform $k\rightarrow 0$. Hence we have $A_2=Y_g$ for arbitrary function $g=g(t)\not=0$. If $g=0$, we have $A_2=W_k$ for arbitrary function $k$ of $t$.
Hence we get essential subalgebras $\mathfrak{L}_{2,4}=<X_1, Y_\psi>$ and $\mathfrak{L}_{2,5}=<X_1, W_\omega>$ for arbitrary functions $\psi\not=0$ and $\omega\not=0$.

(1.3)$A_1=X_{\phi}+Y_1$ for an arbitrary differential function $\phi=\phi(t)$.

Let $A_2=Z_h+X_f+Y_g+W_k$. The condition $[A_1, A_2]=0$ implies two kinds of conditions on $f, h, g$ and $\phi$:

\textbf{Condition 1: } $h=0, \dot{g}=\epsilon(\phi\dot{f}-\dot{\phi}f)$.

\textbf{Condition 2: } $\dot{\phi}=0,h=b\not=0, \dot{g}=\epsilon\phi\dot{f}.$

For the first case, conjugating $A_1$ and $A_2$ by $\exp(\lambda\cdot ad(X_F)+\delta\cdot ad(Y_G))$, one obtains
\begin{eqnarray}
&&A_1\sim A_1+W_{\lambda \overline{F}_0-\delta \overline{G}_0},\,\, \textrm{with}\,\, \overline{G}_0=-\frac{1}{2q}\dot{G},\,\, \overline{F}_0=\frac{1}{2p}(F\dot{\phi}-\dot{F}\phi),\nonumber\\
&&A_2\sim X_f+Y_g+W_{k+\lambda \overline{F}-\delta \overline{G}}, \,\, \textrm{with}\,\, \overline{G}=\frac{1}{2q}(G\dot{g}-\dot{G}g),\,\, \overline{F}=\frac{1}{2p}(F\dot{f}-\dot{F}f).\label{two4}
\end{eqnarray}
Setting
\begin{eqnarray}
\lambda \overline{F}_0-\delta \overline{G}_0=0,\,\,\,
k+\lambda \overline{F}-\delta \overline{G}=0. \label{two3}
\end{eqnarray}
and after tedious calculation, we have the following conclusions about the solutions to system (\ref{two3}).

Case 1: $\dot{g}\not=0$.

The system (\ref{two3}) has infinite number of solutions $F$ and $G$ satisfying
\begin{eqnarray*}
&&G=-\frac{\lambda}{\delta}\epsilon\left((f-\phi g)\dot{F}-(\dot{f}-\dot{\phi}g)F\right)/\dot{g},\\
&&\dot{g}\ddot{F}-\ddot{g}\dot{F}+\epsilon\left((\dot{\phi}\ddot{f}-\dot{f}\ddot{\phi})\right)F=0, \end{eqnarray*}
for $f\not=\phi g$ and
\begin{eqnarray*}
G=\frac{\alpha  \lambda}{\delta}\epsilon F(t),
\end{eqnarray*}
for $f=\phi g$ with $\phi=\alpha$ being a nonzero constant. Taking a solution $F$ and $G$ of the equations as parameters in the above conjugating action (\ref{two4}), we can choose $A_2=X_f+Y_g$ for $\phi, f$ and $g$ satisfying Condition 1. Thus, we have algebra $\mathfrak{L}_{2,6}=<X_{\phi}+Y_1, X_\psi+Y_\Phi>$. For linear independence of $X_{\phi}+Y_1$ and $X_\psi+Y_\Phi$, we need additional condition: $\Phi$ is nonconstant or $\phi$ and $\psi$ are linear independent.

Case 2: $\dot{g}=0$.

By the same procedure given above, we can prove that $<A_1, A_2>\sim <X_{\phi}+Y_1, \alpha X_{\phi}+\beta Y_1>$ for arbitrary functions $\phi$ and arbitrary constants $\alpha$ and $\beta$ with $\alpha\not=\beta$. It is obvious that $<X_{\phi}+Y_1, \alpha X_{\phi}+\beta Y_1>=<X_{\phi}, Y_1>$. While, the last one is equivalent to algebra $\mathfrak{L}_{2,4}$ under adjoint action of $\exp(\delta Z_H)$ for properly selection of parameter $H$ (also see b in Theorem \ref{thh3}). Hence, in this case no new algebra is arisen.

Turning to Condition 2, we can choose $A_2=bZ_1+X_f+Y_{af+c}$  for $\phi=\epsilon a, h=b, g=a f+c$ where $a, b$ and $c$ are arbitrary constants. Now $A_1=Y_1+\epsilon a X_1$.

By Theorem 2, we can take $A_2=Z_1$.  This does not lead to new algebra since the algebra belongs to $\mathfrak{L}_{2,1}$ if $a\not=0$ or $\mathfrak{L}_{2,2}$ if $a=0$.

(1.4)$A_1=W_{\phi}$ for an arbitrary nonzero differential function $\phi=\phi(t)$.

In this case, for $A_2=Z_h+X_f+Y_g+W_k$, the Lie production $[A_1, A_2]=0$ implies
$h\dot{\phi}=0$. Then two cases are arisen.

\textbf{Case 1: } $\phi=a$ is a nonzero constant and $h\not=0$.

\textbf{Case 2: } $h=0$ and $\phi$ is nonzero arbitrary function of $t$.

For the first case, we let $A_1=W_1$. The invariance of $W_k$ under adjoint actions of 2D-CNLS group and adjoint connection between $Z_h+X_f+Y_g+W_k$ and $Z_1$ proved in Theorem \ref{thh2}, here we can take $A_2=Z_1$. This is the same as case $\mathfrak{L}_{2, 3}$.

For the second case, The equivalences among $X_f+Y_g+W_k$, $X_1+Y_{\widetilde{g}}$ and $X_{\widetilde{f}}+Y_1$ when $fg\not=0$ proved in Theorem \ref{thh2} yields algebra $\mathfrak{L}_{2,7}=<W_{\phi}, X_{\psi}+Y_1>$ for arbitray functions $\phi$ and $\psi$. For $g=0, f\not=0$, the algebra $<W_{\phi}, X_1>$ is the same as $\mathfrak{L}_{2,5}$. For the case of $f=0, g\not=0$, we obtain an algebra $<W_{\phi}, Y_1>$ which can be merged into $\mathfrak{L}_{2, 7}$ allowing $\psi=0$.

Obviously, there is an additional algebra $\mathfrak{L}_{2, 8}=<W_{\phi}, W_\psi>$ for $h=f=g=0$ and arbitrary function $\phi\psi\not=0$ and $\phi/\psi\not=$constant.
\subsubsection{Non- Abelian algebras}
(2.1)$A_1=Z_1$.

We take $A_2=Z_h+X_f+Y_g+W_k$. Requiring $[A_1, A_2]=A_1$ and using in Table 1 we find $\dot{h}=1,\dot{f}=\dot{g}=\dot{k}=0$. Hence $A_2=Z_t+bX_1+cY_1+dW_1$. Conjugating by $\exp(2c\cdot ad(Y_1)), \exp(2b\cdot ad( X_1))$ and $\exp(ad (W_{-d\ln t}))$, we can arrive at $A_2=Z_t$ and algebra $\mathfrak{L}_{2, 9}=<Z_1, Z_t>$.

(2.2)$A_1=X_1$.

We take $A_2=Z_h+X_f+Y_g+W_k$. Requiring $[A_1, A_2]=A_1$ and using in Table 1 we find $\dot{h}=2,  \dot{f}=0$. Hence $A_2=Z_{2t+a}+Y_g+W_k$. Conjugating by $\exp(ad(Y_G))$ and $\exp(ad (W_k))$ and appropriately choosing the $G, K$ like doing in Theorem \ref{thh2}, we can arrive at $A_2=Z_{2 t+a}$. Then using Theorem \ref{thh1} with $H=1/2$, one equivalently map $Z_{2t+a}$ to $Z_{t^\prime}$. Hence, we have algebra $\mathfrak{L}_{2,10}=<X_1, Z_{t}>$ after deleting the prime.

(2.3)$A_1=Y_1+X_{\phi}$ for an arbitrary differential function $\phi=\phi(t)$.

Let $A_2=Z_h+X_f+Y_g+W_k$. Then, the condition $[A_1, A_2]=A_1$ implies conditions $h=2t+a, \phi=b, g=\epsilon bf+c$ for arbitrary constants $a, b$ and $c$.

Choosing a nontrivial solutions $F, G$ and $K$ to the following equations
\begin{eqnarray*}
&&4p\lambda \widehat{F}-f=0,\,\,4q\delta \widehat{G}-g=0,\\
&&\delta^2(G\dot{\hat{G}}-\dot{G}\hat{G})-\lambda^2(F\dot{\hat{F}}-\dot{F}\hat{F}) +\lambda\bar{F}-\delta\bar{G}+\nu (2t+a)\dot{K}=0,
\end{eqnarray*}
with $\hat{F}=\frac{1}{4p}(F-(2t+a)\dot{F}),\hat{G}=\frac{1}{4q}(G-(2t+a)\dot{G}), \bar{F}=\frac{1}{2p}(F\dot{f}-\dot{F}f),\bar{G}=\frac{1}{2q}(G\dot{g}-\dot{G}g)$, and  conjugating $A_2$ by $\exp(\delta\cdot ad(Y_G)),\exp(\lambda\cdot ad (X_F))$ and $\exp(\nu\cdot ad (W_K))$ successively we obtain $A_2=Z_{2 t+a}$. It can be proved that the solutions $F$ and $G$ to above equations satisfy $\delta\dot{G}=\epsilon b \lambda \dot{F}$ which guarantees the invariance of $A_1$. Then using the symmetry transformation in Theorem \ref{thh1} with $H=1/2$, we equivalently transform $Z_{2t+a}$ into $Z_t$. Hence we have subalgebra $\mathfrak{L}_{2, 11}=<Y_1+\alpha X_1, Z_{t}>$ for arbitrary constant $\alpha$.

(2.4)$A_1=W_{\phi}$ for an arbitrary differential function $\phi=\phi(t)\not=0$. we take $A_2=Z_h+X_f+Y_g+W_k$. Requiring $[A_1, A_2]=A_1$ yields $\phi+\dot{\phi}h=0$ with $h\not=0$. By theorem \ref{thh2}, we transform $Z_h$ into $Z_1$ and $\phi+\dot{\phi}=0$ in coordinates with prime. Hence we have new algebra $\mathfrak{L}_{2,12}=<W_{e^{-t}}, Z_1>$ after delating the prime.

Summarizing above results, we have the following theorem.
\begin{thh}\label{thh5}
Every two- dimensional subalgebra of {\rm 2D-CNLS} algebra is conjugate under the adjoint group of the {\rm 2D-CNLS} group to precisely one of the following algebras.

1. Abelian algebras:

$\mathfrak{L}_{2,1}=<Z_1, X_1+\alpha Y_1>$, $\forall \alpha \in \mathbb{R}$.

$\mathfrak{L}_{2,2}=<Z_1, Y_1>$.

$\mathfrak{L}_{2,3}=<Z_1, W_1>$.

$\mathfrak{L}_{2,4}=<X_1, Y_\psi>$, $\forall \psi\in C^\infty(R)\backslash\{0\}$.

$\mathfrak{L}_{2,5}=<X_1, W_\omega>$, $\forall \omega\in C^\infty(R)\backslash\{0\}$.

$\mathfrak{L}_{2,6}=<Y_1+X_{\phi}, X_\psi+Y_\Phi>$, either $\Phi$ or $\phi/\psi\not=$constant with $\psi^2+\Phi^2\not=0$ and $\dot{\Phi}=\epsilon(\phi\dot{\psi}-\dot{\phi}\psi)$.

$\mathfrak{L}_{2,7}=<W_{\phi}, X_\psi+Y_1>$, $\forall \phi,\psi\in C^\infty(R), \phi^2+\psi^2 \not=0$.

$\mathfrak{L}_{2,8}=<W_{\phi}, W_\psi>$, $\forall \phi, \psi \in C^\infty(R)$ with $\phi/\psi\not=$constant.

2. Non-Abelian algebras satisfying $[A_1, A_2]=A_1$:

$\mathfrak{L}_{2, 9}=<Z_1, Z_t>$.

$\mathfrak{L}_{2,10}=<X_1, Z_{t}>$.

$\mathfrak{L}_{2,11}=<Y_1+\alpha X_1, Z_{t}>$, $\forall \alpha\in \mathbb{R}$.

$\mathfrak{L}_{2,12}=<W_{e^{-t}}, Z_1>$.
\end{thh}
\subsection{Three-dimensional subalgebras}

A real three- dimensional Lie algebra can be either simple or solvable. We will consider these two cases separately.

\textbf{1. Simple Lie algebras}

By Lemma 3, this algebra has a two- dimensional non- Abelian subalgebra with commutation relations (\ref{sl}). We will identify $\{A_1, A_2\}$ with one of the algebras in $\mathfrak{L}_{2,9}\sim\mathfrak{L}_{2,12}$, i.e., consider it to be already in standard form.

Let us start with

$\mathfrak{L}_{2,9}: A_1=Z_1, A_2=Z_t$, $A_3=Z_h+X_f+Y_g+W_k$.

Imposing (\ref{sl}) we find $A_3=Z_{t^2}$.

Then consider

$\mathfrak{L}_{2,10}: A_1=X_1, A_2=Z_{t}$, $A_3=Z_h+X_f+Y_g+W_k$;

$\mathfrak{L}_{2,11}: A_1=Y_1+\alpha X_1, A_2=Z_{t}$, $A_3=Z_h+X_f+Y_g+W_k$;

$\mathfrak{L}_{2,12}: A_1=W_{e^{-t}}, A_2=Z_1$, $A_3=Z_h+X_f+Y_g+W_k$.

It is easy to see that $[A_1, A_3]=2A_2$ cannot be satisfied for all the three cases.

We thus have reobtained  algebras $\mathfrak{L}_s$ spanned by $\{Z_1, Z_t, Z_{t^2}\}$ given in (\ref{sp}). This is unique three dimensional simple algebra admitted by 2D-CNLS system (\ref{e1}).

\textbf{2. Solvable Lie algebras}

A solvable three dimensional Lie algebra always have a two- dimensional Abelian ideal. We assume that the ideal spanned by $\{A_1, A_2\}$ is already in standard form $\mathfrak{L}_{2,1}\sim\mathfrak{L}_{2,8}$ and look for a third element $A_3=Z_h+X_f+Y_g+W_k$ so that $A_1, A_2$ and $A_3$ span the representative of the class of subalgebras.

We first use the commutation relations, then simplify $A_3$ using the stabilizer (isotropy) group of $\{A_1, A_2\}$ in the invariance group of the 2D-CNLS equations.

Let us now realize this procedure for each two- dimensional Abelian subalgebras in $\mathfrak{L}_{2,1}\sim\mathfrak{L}_{2,8}$.

a. $\mathfrak{L}_{2,1}=<Z_1, X_1+\alpha Y_1>$ for arbitrary constant $\alpha$.

Imposing (\ref{three}) and using the commutator in Table 1, we obtain
\begin{eqnarray}\dot{h}=a=2d,\,\, \dot{f}=b, \dot{g}=\alpha b, \dot{k}=0, c=0, (\alpha^2-\epsilon)b=0.\label{2.1}\end{eqnarray}
Naturally, two cases are arisen.

\textbf{Case $a\not=0$}. In this case, according to the conditions (\ref{2.1}), we can set $A_3=aZ_t+b(X_t+\alpha Y_t)+c_1Y_1+c_2W_1$. Conjugating the $A_3$ by $\exp(\delta \cdot ad(Y_1))$ and $\exp(\mu \cdot ad(W_1))$ with selections of $\delta$ and $\mu$ satisfying $\frac{\delta}{2}-c_2=0$ and $c_2-\frac{\alpha b \delta}{2q}+\mu=0$, we can arrive $A_3\sim Z_t+b(X_t+\alpha Y_t)$. For $b\not=0$, then, conjugating further by $\exp(-2\alpha b \cdot ad(Y_t))$ and $\exp(\-2\alpha b \cdot ad(X_t))$, we have $A_3\sim Z_{2t}$. Hence we have algebra $\mathfrak{L}_{3,1}=<Z_1, X_1+\alpha Y_1, Z_{2t}>$. Here $M=\left[
  \begin{array}{cc}
    1 & 0 \\
    0 & 1\\
  \end{array}\right]$
for arbitrary constant $\alpha$.

\textbf{Case $a=0$}. If $b=0$, then this algebra is Abelian $\mathfrak{L}^0_{3,1}=<Z_1,X_1+\alpha Y_1, \beta X_1+\gamma Y_1+\mu W_1>$ for arbitrary constants $\alpha, \beta, \gamma, $ and $\mu$. If $b\not=0$ and after scaling, we set $b=1$ and $A_3=X_t+\alpha Y_t+c_1Y_1+c_2 W_1$ for arbitrary constants $c_1, c_2$ and $\alpha^2=\epsilon$. Conjugating the $A_3$ through $\exp(\delta\cdot ad(Y_1))$ by selecting $\delta=2qc_2/\alpha$, we can remove the term of $W_1$ from $A_3$. Hence we have Lie algebra $\mathfrak{L}_{3,2}=<Z_1, X_1+\alpha Y_1, X_t+\alpha Y_t+\beta Y_1>$ for $\alpha$ satisfying $\alpha^2=\epsilon$ and arbitrary constant $\beta$. Here $M=\left[
  \begin{array}{cc}
    0 & 1\\
    0 & 0\\
  \end{array}\right].$

b. $\mathfrak{L}_{2,2}=<Z_1, Y_1>$.

Imposing (\ref{three}) and using the commutators in Table 1, we obtain
\begin{eqnarray}b=c=0, \dot{h}=a=2d,\,\, \dot{f}=\dot{g}=\dot{k}=0.\label{2.2}\end{eqnarray}
Hence we can write $A_3=a Z_t+c_1X_1+c_2 W_1$ for arbitrary constants $c_1$ and $c_2$.
If $a=0$, then this is an Abelian algebra $\mathfrak{L}^0_{3,2}=<Z_1, Y_1, \alpha X_1+\beta W_1>$ for arbitrary not all zero constants $\alpha$ and $\beta$. If $a\not=0$, scaling and conjugating through $\exp(\lambda\cdot ad(X_1))$ with $\lambda=2c_1/a$, we have $A_3\sim Z_t+c_2 W_1$ for arbitrary constant $c_2$. Hence we have Lie algebra $\mathfrak{L}_{3,3}=<Z_1, Y_1, Z_{2t}+\alpha W_1>$ for arbitrary constant $\alpha$ and corresponding matrix $M=\left[
  \begin{array}{cc}
    1 & 0\\
    0 & 1\\
  \end{array}\right].$

c. $\mathfrak{L}_{2,3}=<Z_1, W_1>$.

Imposing (\ref{three}) and using the commutators in Table 1, we obtain
\begin{eqnarray}d=c=0, \dot{h}=a,\,\, \dot{f}=\dot{g}=0,\,\, \dot{k}=b.\label{2.3}\end{eqnarray}
Hence we can write $A_3=a Z_t+c_1X_1+c_2 Y_1+bW_t$ for arbitrary constants $c_1$ and $c_2$. Similar to the above cases, if $a=0$, we have an Abelian algebra $\mathfrak{L}_{3,3}^0=<Z_1,W_1,\alpha X_1+\beta Y_1>$ for arbitrary not all zero constants $\alpha$ and $\beta$. If $a\not=0$, applying $\exp(\delta \cdot ad(Y_1))$ and $\exp(\lambda\cdot ad(X_1))$ on the $A_3$ for suitable selections of $\delta$ and $\lambda$, we can remove the $X_1$ and $Y_1$ terms from the $A_3$ and have  $A_3\sim Z_t+\alpha W_t$ for arbitrary constant $\alpha$. Further, conjugating by $\exp(-\alpha W_t)$, we make $A_3\sim Z_t$; If $a=0$, then either $c_1\not=0$ or $c_2\not =0$ including $b=0$, we always make $A_3\sim c_1X_1+c_2Y_1$ by applying $\exp(\lambda\cdot ad(Y_{t^2})$ or $\exp(\lambda\cdot ad(X_{t^2})$. This leads to Abelian algebras $\mathfrak{L}_{3,3}^0$. Hence we only left the case $c_2=c_3=0$ and $b\not=0$. In this case $A_3\sim W_t$.

Consequently, we have algebras
$\mathfrak{L}_{3,4}=<Z_1, W_1, Z_t>$ and $\mathfrak{L}_{3,5}=<Z_1, W_1, W_t>$ with corresponding matrix $M=\left[
  \begin{array}{cc}
    1 & 0\\
    0 & 0\\
  \end{array}\right]$
and $M=\left[
  \begin{array}{cc}
    0 & 1\\
    0 & 0\\
  \end{array}\right],$ respectively.

d. $\mathfrak{L}_{2,4}=<X_1, Y_{\psi}>$ for arbitrary function $\psi$ of $t$.

Imposing (\ref{three}) and using the commutators in Table 1, we obtain
\begin{eqnarray}b=c=0, \dot{h}=2a,\,\, \dot{f}=0,\,\, (a-d)\psi=h\dot{\psi},\,\, \psi\dot{g}-\dot{\psi}g=0.\,\,\label{2.4}\end{eqnarray}
The last equation implies $g=\widetilde{c}\psi$ for arbitrary constant $\widetilde{c}$. Hence we can write $A_3=Z_h+W_k$.

If $h\not=0$, conjugating through $\exp(ad (W_K))$, we can remove $W_k$ term from $A_3$ by selecting $K$ satisfying $h\dot{K}+k=0$. Hence $A_3\sim Z_h$. From (\ref{2.4}), we have $\psi=\widetilde{c}h^{\frac{a-d}{2a}}$ for $a\not=0$ and $\psi=\widetilde{c}\exp(-\frac{d}{h})t$ for $a=0$. Here $\widetilde{c}$ denotes an arbitrary constant. For the both cases of $\psi$, we use Theorem \ref{thh3} with translating and scaling variable $t$, we get algebras $\mathfrak{L}_{3,6}=<X_1, Y_{(2t)^\alpha}, Z_{2t}>$ for arbitrary constant $\alpha$ and $\mathfrak{L}_{3,7}=<X_1, Y_{e^{-t}}, Z_1>$. They have matrices $M=\left[
\begin{array}{cc}
    1 & 0\\
    0 & 1-\alpha\\
\end{array}\right]$ and $M=\left[
\begin{array}{cc}
    0 & 0\\
    0 & 1\\
\end{array}\right]$, respectively.

If $h=0$, we get Abelian algebra $\mathfrak{L}_{3,4}^0=<X_1, Y_\psi, W_\omega>$ for arbitrary nonzero functions $\psi$ and $\omega$.

e. $\mathfrak{L}_{2,5}=<X_1, W_\omega>$ for arbitrary function $\omega$ of $t$.

Imposing (\ref{three}) and using the commutators in Table 1, we obtain
\begin{eqnarray}c=0, \dot{h}=2a,\,\, \dot{f}=-2pb\omega,\,\, h\dot{\omega}=-d\omega.\label{2.5}\end{eqnarray}

For simplification of $A_3$, three cases are considered.

\textbf{Case $\dot{h}\not=0$.}

Applying $\exp(\delta \cdot ad(Y_G))$ and letting $G$ be any solution of equation $g-\delta(\frac{1}{2}G\dot{h}-\dot{G}h)=0$, we remove the $Y_g$ term from $A_3$, i.e., $A_3\sim A_3^\prime=Z_h+X_f+W_{\widetilde{k}}$ for some function $\widetilde{k}$ depending on $G$ and $k$. Now applying $\exp(\lambda\cdot ad(X_F))$ on $A_3^\prime$ and letting $F$ satisfy
\begin{eqnarray}
f-\lambda(\frac{1}{2}F\dot{h}-\dot{F}h)=0.\label{F2.50}
\end{eqnarray}

Meantime, in order to keep the invariance of algebra $\mathfrak{L}_{2,5}$, we also let
\begin{eqnarray}
\dot{F}=2p\rho \omega,\label{F2.51}
\end{eqnarray}
for some constant $\rho$.

Under the condition $a\not=d$, there exist solutions $F$ and $\rho=b/(a-d)$ to equations (\ref{F2.50}) and (\ref{F2.51}). Therefore, under $a\not=d$, we reach $A_3\sim Z_h+W_{\widehat{k}}$ for another function $\widehat{k}$ depending on $G, F$ and $\rho$. Finally applying $\exp(\mu\cdot ad(W_K))$, we can remove the $W$ term from $A_3$ by requiring $h\dot{K}+\widehat{k}=0$, i.e., $A_3\sim Z_h.$

Solving equations (\ref{2.5}), we have $\omega=\widetilde{c} h^{-\frac{d}{2a}}$ for arbitrary constant $\widetilde{c}$.

Consequently, after using the scaling and translation for the variable $t$, we obtain equivalent algebra $\mathfrak{L}_{3,8}=<X_1, W_{(2t)^\nu}, Z_{2t}>$ with $\nu\not=-\frac{1}{2}$. It corresponds to matrix $M=\left[
\begin{array}{cc}
    1& 0\\
    0 & -2\nu\\
\end{array}\right]$.

Similarly, when $a=d$, we obtain equivalence algebra $\mathfrak{L}_{3,9}=<X_1, W_{1/\sqrt{2t}}, Z_{2t}+4p \alpha X_{\sqrt{2t}}>$ with matrix $M=\left[
\begin{array}{cc}
    1&0\\
    0 & 1\\
\end{array}\right]$ for arbitrary constant $\alpha$.

\textbf{Case $\dot{h}=0, h\not=0$}. We obtain algebra $\mathfrak{L}_{3, 10}=<X_1, W_{e^{-t}}, Z_1>$ which corresponds to matrix $M=\left[
\begin{array}{cc}
    0& 0\\
    0 & 1\\
\end{array}\right]$.

\textbf{Case $h=0$.}  Conjugating $A_3$ through $\exp(\delta\cdot ad(Y_G))$, we can remove the term $W_k$ from it. Hence we obtain algebra $\mathfrak{L}_{3, 11}=<X_1, W_{\omega}, X_f+Y_\psi>$ with $\dot{f}=-2p \alpha^2\omega$ and matrix $M=\left[
\begin{array}{cc}
    0& 1\\
    0 & 0\\
\end{array}\right]$ for arbitrary functions $\omega, \psi$ and constant $\alpha\not=0$.

If $\alpha=0$, we have algebra $\mathfrak{L}_{3,4}^0$ again.

f. $\mathfrak{L}_{2,6}=<Y_1+X_{\phi}, X_\psi+Y_\Phi>$ for arbitrary functions $\phi, \psi$ and $\Phi$ satisfying $\dot{\Phi}=\epsilon(\phi\dot{\psi}-\dot{\phi}\psi)$.

Let $A_1=X_{\phi}+Y_1, A_2=X_\psi+Y_\Phi$ and $A_3=Z_h+X_f+Y_g+W_k$.  Imposing (\ref{three}) and noticing Abelian property of $A_1$ and $A_2$, we obtain
\begin{eqnarray*}
\hspace*{1cm}\textrm{(I)}\,\,\left\{\begin{array}{ll}
     \frac{1}{2}\phi\dot{h}-\dot{\phi}h=a\phi+b\psi&\\
     \frac{1}{2}\dot{h}=a+b\Phi&\\
     \dot{g}=\epsilon(\phi\dot{f}-\dot{\phi}f)&
     \end{array}\right.\,\,\,
\textrm{(II)}\,\,\left\{\begin{array}{ll}
     \frac{1}{2}\psi\dot{h}-\dot{\psi}h=c\phi+d\psi&\\
     \frac{1}{2}\Phi\dot{h}-h\dot{\Phi}=c+d\Phi&\\
     \Phi\dot{g}-\dot{\Phi}g=\epsilon(\psi\dot{f}-\dot{\psi}f)&
     \end{array}\right.\,\,\,
\textrm{(III)}\,\,\begin{array}{ll}
     \dot{\Phi}=\epsilon(\phi\dot{\psi}-\dot{\phi}\psi)&
     \end{array}
\end{eqnarray*}
Particularly, for the linear combinations of $A_1$ and $A_2$: $B_1=\alpha A_1+\beta A_2$ and $B_2=\gamma A_1+\sigma A_2$ with $\alpha\gamma-\beta\sigma\not=0$, we can transform $B_1$ into $A_1$ and $B_2$ into $A_2$ in different functions $\phi,\psi$ and $\Phi$ by $\exp(\mu\cdot ad(Z_H))$. Thus we assume that $M$ is already in standard form (see Lemma 3).

(f.1)\,\,$M=\left[\begin{array}{cc}
    0&0\\
    0&0\\
\end{array}\right].$

Corresponding to the cases $h\not=0$ and $h=0$, we obtain two algebras, namely $<X_1, Y_1, Z_1>$ which belongs to $\mathfrak{L}_{3,2}^0$ when $\beta=0$ and $\mathfrak{L}_{3,5}^0=<X_\phi+Y_1, X_\psi+Y_\Phi, X_f+Y_g+W_k>$ for arbitrary solutions  $\phi, \psi, f, g$ to (I)$_3$, (II)$_3 $ and (III) with either $\{\phi, \psi\}$ is linear independent or $\Phi$ is not constant, where $k$ is an arbitrary function of $t$.

(f.2)\,\,$M=\left[\begin{array}{cc}
    1&0\\
    0&0\\
\end{array}\right].$

Solving (I)-(II) with this matrix $M$, we obtain two algebras $\mathfrak{L}_{3, 12}=<\beta X_1+Y_1, X_{\sqrt{2t}}+\epsilon\beta Y_{\sqrt{2t}}, Z_{2t}>$ for $\epsilon \beta^2\not=1$ and $\mathfrak{L}_{3, 13}=<\beta X_1+Y_1, \beta X_{\sqrt{2t}}+Y_{\sqrt{2t}}, Z_{2t}+\beta X_g+Y_g>$ for $\epsilon \beta^2=1$ and arbitrary function $g$ of $t$.

(f.3)\,\,$M=\left[\begin{array}{cc}
    1&0\\
    0&\alpha\\
\end{array}\right].$

Solving (I)-(II) with this matrix $M$, we obtain three algebras: $\mathfrak{L}_{3, 14}=<\beta X_1+Y_1, X_{{(2t)}^\frac{1-\alpha}{2}}+\epsilon\beta Y_{{(2t)}^\frac{1-\alpha}{2}},$ $Z_{2t}>$ for both $\alpha\not=1$ and $\epsilon \beta^2\not=1$; $\mathfrak{L}_{3, 15}=<\beta X_1+Y_1, X_{{(2t)}^\frac{1-\alpha}{2}}+\epsilon\beta Y_{{(2t)}^\frac{1-\alpha}{2}}, Z_{2t}+\beta X_g+Y_g>$ for $\alpha\not=1$ and $\epsilon \beta^2=1$ and arbitrary function $g$ of $t$; $\mathfrak{L}_{3, 16}=<X_1,Y_1, Z_{2t}>$ for $\alpha=1$.

(f.4)\,\,$M=\left[\begin{array}{cc}
    \alpha& 1\\
    -1&\alpha\\
\end{array}\right].$

In this case, from (I)-(II) we see that we have to $h\not=0$. In order to solve system (I)-(III), we use a way differ frpm above procedure. At first, we  perform a conjugating by $\exp(\mu\cdot ad(Z_H))$ changing the basis $A_1, A_2$ and $A_3$ into $B_1=X_\phi+Y_\Psi$, $B_2=X_\psi+Y_\Phi$ and $B_3=Z_1+X_f+Y_g+W_k$, respectively (without loss of generality, we take same notations on these group labeled functions). The conditions (\ref{three})  now implies
\begin{eqnarray*}
\hspace*{0.5cm}\textrm{(I)}^\prime \left\{\begin{array}{l}
     -\dot{\phi}=\alpha\phi+\psi,\\
     -\dot{\Psi}=\alpha\Psi+\Phi,\\
     \Psi\dot{g}-\dot{\Psi}g=\epsilon(\phi\dot{f}-\dot{\phi}f).
     \end{array}\right.
\textrm{(II)}^\prime \left\{\begin{array}{l}
     -\dot{\psi}=\alpha\psi-\phi,\\
     -\dot{\Phi}=\alpha\Phi-\Psi,\\
     \Phi\dot{g}-\dot{\Phi}g=\epsilon(\psi\dot{f}-\dot{\psi}f).
     \end{array}\right.
\textrm{(III)}^\prime \begin{array}{l}
     \Psi\dot{\Phi}-\dot{\Psi}\Phi=\epsilon(\phi\dot{\psi}-\dot{\phi}\psi).
     \end{array}
\end{eqnarray*}

Solving first two equations in (I)$^\prime_{1-2}$ and (II)$^\prime_{1-2}$ respectively, we obtain
\begin{eqnarray*}
&&\phi=e^{-\alpha t}(c_1\cos t+c_2\sin t);\,\,\, \psi=e^{-\alpha t}(-c_2\cos t+ c_1\sin t);\\
&&\Phi=e^{-\alpha t}(c_3\cos t+c_4\sin t);\,\,\, \Psi=e^{-\alpha t}(c_4\cos t-c_3\sin t),
\end{eqnarray*}
for arbitrary constants $c_1\sim c_4$ satisfying $\epsilon(c_1^2+c_2^2)=(c_3^2+c_4^2)$ from (III)$^\prime$, which implies that $\phi=\psi=0$ is equivalent to  $\Psi=\Phi=0$.

Hence, we have $c_1^2+c_2^2\not=0$. Under this condition, we always choice a basis of algebra $\mathfrak{L}_{2,6}=<A_1, A_2>$ through linear combinations of $A_1$ and $A_2$ such that $c_1=1$ and $c_2=0$, which leads to $\phi=e^{-\alpha t}\cos t, \psi=e^{-\alpha t} \sin t$ and $\Phi=\rho\phi+\beta\psi, \Psi=\beta\phi-\rho\psi$ for arbitrary constants $\rho$ and $\beta$ satisfying  $\epsilon=\rho^2+\beta^2$.

Solving (I)$^\prime_3$ and (II)$^\prime_3$, we obtain
\begin{eqnarray}
&&f(t)=\delta (\rho \phi-\beta\psi)+\sigma \psi,\,\, g(t)=\sigma(\rho \phi+\beta\psi)-\delta \epsilon \psi.\label{fg1}
\end{eqnarray}
if $\rho\not=0$. Here $\delta, \sigma, \rho$ and $\beta$ are arbitrary constants with satisfying $\epsilon=\rho^2+\beta^2$; and
\begin{eqnarray}
&&g(t)=\beta f(t) \label{fg2}
\end{eqnarray}
if $\rho=0$. Here $f$ is arbitrary function of $t$ and $\epsilon=\beta^2$.

Hence we obtain algebra $\mathfrak{L}_{3, 17}$ and $\mathfrak{L}_{3, 18}$ spanned by $\{A_1, A_2, A_3\}$ with $A_1=X_{\phi}+Y_\Psi, A_2=X_\psi+Y_\Phi$ and $A_3=Z_1+X_f+X_g$ in which $f$ and $g$ are given by (\ref{fg1}) and (\ref{fg2}), respectively.

(f.5)\,\,$M=\left[\begin{array}{cc}
    0& 0\\
    1& 0\\
\end{array}\right].$

In this case, we obtain two algebras $\mathfrak{L}_{3, 19}$ spanned by $A_1=\beta X_1+Y_1, A_2=X_{\beta t+\nu}+Y_t$ and $A_3=-Z_1$ for arbitrary constant $\nu$ and $\mathfrak{L}_{3, 20}$ spanned by $A_1=\beta X_1+Y_1, A_2=\beta X_t+Y_t$ and $A_3=-Z_1+\beta X_g+Y_g$ for arbitrary function $g$ of $t$. In both cases, $\beta$ satisfies  $\epsilon\beta^2=1$.

(f.6)$\,\,M=\left[\begin{array}{cc}
    1& 0\\
    1& 1\\
\end{array}\right].$

We obtain two algebras: $\mathfrak{L}_{3,21}$ spanned by $A_1=\beta X_1+Y_1, A_2=X_{-\frac{\beta}{2} \ln t+\nu}+Y_{-\frac{1}{2}\ln t+\sigma}$ and $A_3=Z_{2t}$ for arbitrary constants $\nu$ and $\sigma$; and $\mathfrak{L}_{3, 22}$ spanned by $A_1=\beta X_1+Y_1, A_2=X_{-\frac{\beta}{2} \ln t}+Y_{-\frac{1}{2}\ln t}$ and $A_3=Z_{2t}+\beta X_g+Y_g$ for arbitrary function $g$ of $t$. In both cases, $\beta$ satisfies  $\epsilon\beta^2=1$.

g. $\mathfrak{L}_{2,7}=<W_{\phi}, X_\psi+Y_1>$ for arbitrary functions $\phi$ and $\psi$.

Corresponding to each case of standard matrices $M$, we obtain seven algebras.

(g.1)\,\, $M=\left[\begin{array}{cc}
    0& 0\\
    0& 0\\
\end{array}\right].$ In this case we have two algebras.
Algebra $\mathfrak{L}_{3,6}^0$ is spanned by $A_1=W_1, A_2=\beta X_1+Y_1$ and $A_3=Z_1$ for arbitrary constant $\beta$; Algebra $\mathfrak{L}_{3, 7}^0$ is spanned by $A_1=W_\phi, A_2=X_\psi+Y_1$ and $A_3=X_f+Y_g$ for arbitrary functions $f, g,\phi$ with $f^2+g^2\not=0$ and satisfying $\dot{g}=\epsilon(\psi\dot{f}-\dot{\psi}f)$. If $f=g=0$, we have an Abelian algebra $\mathfrak{L}_{3,7}^\prime=<W_\phi, X_\psi+Y_1, W_k>$ where arbitrary functions $\phi$ and $k$ are linear independent.

(g.2)\,\, $M=\left[\begin{array}{cc}
    1& 0\\
    0& 0\\
\end{array}\right].$ In this case we have algebra $\mathfrak{L}_{3, 23}$ spanned by $A_1=W_{e^{-t}}, A_2=\beta X_1+Y_1+\alpha W_1$ and $A_3=Z_1$ for arbitrary constants $\alpha, \beta$;

(g.3)\,\, $M=\left[\begin{array}{cc}
    0& 0\\
    1& 0\\
\end{array}\right].$ In this case we have two algebras. Algebra $\mathfrak{L}_{3, 24}$ spanned by $A_1=W_1, A_2=\beta X_1+Y_1+ W_t$ and $A_3=Z_1$ for arbitrary constant $\beta$; Algebra $\mathfrak{L}_{3, 25}$ spanned by $A_1=W_\phi, A_2=X_\psi+Y_1$ and $A_3=X_f+Y_g$ for arbitrary functions $f, g, \phi$ and $\psi$ satisfying $\dot{g}=2q\phi+\epsilon({\psi\dot{f}- f\dot{\psi}})$; $\mathfrak{L}_{3,26}=<W_\phi, Y_1, Y_g+W_k>$ for arbitrary functions $g, k$ and $\phi\not=0$  satisfying $\dot{g}=2q\phi$;

(g.4)\,\, $M=\left[\begin{array}{cc}
    1& 0\\
    0& d\\
\end{array}\right].$ In this case we have algebra $\mathfrak{L}_{3, 27}=<W_{t^{-\nu}}, \beta X_1+Y_1, Z_{\nu^{-1}t}>$ for arbitrary  constants $\beta$ and $d=(2\nu)^{-1}\not=0$.

(g.5)\,\, $M=\left[\begin{array}{cc}
    1& 0\\
    1& 1\\
\end{array}\right].$ In this case we have no new algebra;

k. $\mathfrak{L}_{2,8}=<W_{\phi}, W_\psi>$ for arbitrary nonzero functions $\phi$ and $\psi$ with $\phi/\psi\not=$constant.

We consider two cases $\h\not=0$ and $h=0$ separately.

\textbf{Case: $h\not=0$.} According to the Theorem \ref{thh1} and adjoint action Table 4, we can take  $A_3=Z_h$  without changing $<A_1=W_{\phi}, A_2=W_\psi>$. Conjugating by $\exp(\mu\cdot ad(Z_H))$ and changing the basis $A_1, A_2$ and $A_3$ into $B_1=X_\phi+Y_\Psi$, $B_2=X_\psi+Y_\Phi$ and $B_3=Z_1+X_f+Y_g+W_k$, respectively (without loss of generality, we take same notations on these group labeled functions). Thus the conditions (\ref{three}) implies
\begin{eqnarray}
     -\dot{\phi}=a\phi+b\psi,\,\,\, -\dot{\psi}=c\psi+d\phi.\label{l28}
\end{eqnarray}
Corresponding to each case of standard matrices $M$, we solve equations (\ref{l28}) and  obtain following algebras.

(k.1)\,\, $M=\left[\begin{array}{cc}
    1& 0\\
    0& 0\\
\end{array}\right].$ Algebra $\mathfrak{L}_{3, 28}=<W_{e^{-t}}, W_1, Z_1>$.

(k.2)\,\, $M=\left[\begin{array}{cc}
    0& 0\\
    1& 0\\
\end{array}\right].$ Algebra $\mathfrak{L}_{3, 29}=<W_{1}, W_{-t}, Z_1>$.

(k.3)\,\, $M=\left[\begin{array}{cc}
    1& 0\\
    0& d\\
\end{array}\right].$ Algebra $\mathfrak{L}_{3, 30}=<W_{e^{-t}}, W_{e^{-dt}}, Z_1>$, $\forall d\in \mathbb{R}\backslash\{0\}$.

(k.4)\,\, $M=\left[\begin{array}{cc}
    a& 1\\
    -1& a\\
\end{array}\right].$ Algebra $\mathfrak{L}_{3, 31}=<W_{e^{-at}\cos t}, W_{e^{-at}\sin t}, Z_{1}>$, $\forall a\in \mathbb{R}$.

(k.5)\,\, $M=\left[\begin{array}{cc}
    1& 0\\
    1& 1\\
\end{array}\right].$ Algebra $\mathfrak{L}_{3, 32}=<W_{e^{t}}, W_{te^{t}}, Z_{-1}>$.

\textbf{Case: $h=0$.} In this case, we have only Abelian algebra $\mathfrak{L}_{3, 8}^0=<W_\phi, W_\psi, X_f+Y_g>$ for arbitrary functions $\phi, \psi, f$ and $g$ with satisfying $\phi/\psi\not=$ constant and $f^2+g^2\not=0$. Here we notice that  algebra $\mathfrak{L}_{3,7}^\prime$ is contained in algebra $\mathfrak{L}_{3, 8}^0$.

\textbf{Remark 2: }The algebras $\mathfrak{L}_{3,2}, \mathfrak{L}_{3,13}-\mathfrak{L}_{3,22}$ only exist for $\epsilon>0$. This shows the effect of the physical constants $p$ and $q$ on the extending of symmetries of system (\ref{e1}) when they have the same sign.

In summarizing this section, we present a list of representatives of all equivalent classes of three dimensional subalgebras of 2D-CNLS algebra by specifying the matrix $M$ in (\ref{stand}). The solvable algebras are given in order $\{A_1, A_2, A_3\}$ and $\{A_1, A_2\}$ is Abelian ideal and the action of $A_3$ is given in (\ref{three}).
\begin{thh}\label{thh6}
Every three- dimensional subalgebra of {\rm 2D-CNLS} algebra is conjugate under the adjoint group of the {\rm 2D-CNLS} group to precisely one of the following algebras of lists 1-7.
\end{thh}
1. Abelian algebras  with $M=\left[\begin{array}{cc}
    0& 0\\
    0& 0\\
\end{array}\right]:$

$\mathfrak{L}^0_{3,1}=<Z_1,X_1+\alpha Y_1, \beta X_1+\gamma Y_1+\mu W_1>$, $\forall \alpha, \beta, \gamma, \mu\in \mathbb{R}$.

$\mathfrak{L}^0_{3,2}=<Z_1, Y_1, \alpha X_1+\beta W_1>$, $\forall \alpha, \beta \in R, \alpha^2+\beta^2\not=0$.

$\mathfrak{L}_{3,3}^0=<Z_1, W_1, \alpha X_1+\beta Y_1>$, $\forall \alpha, \beta\in R, \alpha^2+\beta^2\not=0$.

$\mathfrak{L}_{3,4}^0=<X_1, Y_\psi, W_\omega>$, $\forall \psi,\omega \in C^\infty(R)$.

$\mathfrak{L}_{3,5}^0=<X_\phi+Y_1, X_\psi+Y_\Phi, X_f+Y_g+W_k>, \forall\textrm{ solutions }  \phi, \psi, \Phi, f, g\in C^\infty(\mathbb{R}) $ to (I)$_3$, (II)$_3 $ and (III) with either $\{\phi, \psi\}$ is linear independent or $\Phi$ is not constant.

$\mathfrak{L}_{3,6}^0=<W_1, \beta X_1+Y_1, Z_1>$, $\forall \beta \in \mathbb{R}$.

$\mathfrak{L}_{3, 7}^0=<W_\phi, X_\psi+Y_1, X_f+Y_g>$, $\forall \phi,\psi, f, g\in C^\infty(\mathbb{R})$ satisfying $f^2+g^2\not=0$ and $\dot{g}=\epsilon(\psi\dot{f}-\dot{\psi}f)$.

$\mathfrak{L}_{3, 8}^0=<W_\phi, W_\psi, X_f+Y_g>$, $\forall \phi, \psi, f, g \in C^\infty(\mathbb{R})$ with $\phi/\psi\not=$ constant and $f^2+g^2\not=0$.

2. Decomposable, non-Abelian  algebras, with $M=\left[\begin{array}{cc}
    1& 0\\
    0& 0\\
\end{array}\right]:$

$\mathfrak{L}_{3,4}=<Z_1, W_1, Z_t>$.

$\mathfrak{L}_{3,7}=<Y_{e^{-t}}, X_1, Z_1>$.

$\mathfrak{L}_{3, 10}=<W_{e^{-t}}, X_1, Z_1>$.

$\mathfrak{L}_{3, 12}=<\beta X_1+Y_1, X_{\sqrt{2t}}+\epsilon\beta Y_{\sqrt{2t}}, Z_{2t}>$, $\textrm{ for }\epsilon \beta^2\not=1$.

$\mathfrak{L}_{3, 13}=<\beta X_1+Y_1, \beta X_{\sqrt{2t}}+Y_{\sqrt{2t}}, Z_{2t}+\beta X_g+Y_g>$, for $\epsilon \beta^2=1,\forall g\in C^\infty (\mathbb{R})$.

$\mathfrak{L}_{3, 23}=<W_{e^{-t}}, \beta X_1+Y_1+\alpha W_1, Z_1>$, $\forall \alpha, \beta\in \mathbb{R}$.

$\mathfrak{L}_{3, 28}=<W_{e^{-t}}, W_1, Z_1>$.

3. Nilpotent algebras with $M=\left[\begin{array}{cc}
    0& 0\\
    1& 0\\
\end{array}\right]:$

$\mathfrak{L}_{3,2}=<X_1+\alpha Y_1, Z_1, X_t+\alpha Y_t+\beta Y_1>$, $\alpha^2=\epsilon, \forall\beta\in \mathbb{R}$.

$\mathfrak{L}_{3,5}=< W_1, Z_1, W_t>$.

$\mathfrak{L}_{3, 11}=< W_{\omega}, X_1, X_f+Y_\psi>, \dot{f}=-2p \alpha^2\omega,  \forall \omega, \psi\in C^\infty(R), \alpha\in R\backslash\{0\}$.

$\mathfrak{L}_{3, 19}<\beta X_1+Y_1, X_{\beta t+\nu}+Y_t, A_3=-Z_1>$, $\forall \nu\in \mathbb{R}$ and $\epsilon\beta^2=1$.

$\mathfrak{L}_{3, 20}=<\beta X_1+Y_1, \beta X_t+Y_t,-Z_1+\beta X_g+Y_g>$, $\forall g\in C^\infty(\mathbb{R})$ and $\epsilon\beta^2=1$.

$\mathfrak{L}_{3, 24}=<W_1, A_2=\beta X_1+Y_1+W_t,Z_1>$, $\forall \beta \in \mathbb{R}$.

$\mathfrak{L}_{3, 25}=<W_\phi, X_\psi+Y_1, X_f+Y_g>$, $\forall f, g, \phi, \psi\in C^\infty (\mathbb{R})$ satisfying $\dot{g}=2\alpha q\phi+\epsilon({\psi\dot{f}- f\dot{\psi}})$, $\alpha\in R\backslash \{0\}$.

$\mathfrak{L}_{3,26}=<W_\phi, Y_1, Y_g+W_k>$, $\forall g, k, \phi \in C^\infty (R)$  satisfying $\dot{g}=2q\phi$.

$\mathfrak{L}_{3, 29}=<W_{1}, W_{-t}, Z_1>$.

4. Diagonal action on ideal with $M=\left[\begin{array}{cc}
    1& 0\\
    0& a\\
\end{array}\right], a\not=0:$

$\mathfrak{L}_{3,1}=<Z_1, X_1+\alpha Y_1, Z_{2t}>, a=1, \forall \alpha \in \mathbb{R}$.

$\mathfrak{L}_{3,3}=<Z_1, Y_1, Z_{2t}+\alpha W_1>$, $\forall\alpha \in \mathbb{R}, a=1$.

$\mathfrak{L}_{3,6}=<X_1, Y_{(2t)^\alpha}, Z_{2t}>$, $\forall \alpha\in \mathbb{R}, a=1-\alpha$.

$\mathfrak{L}_{3,8}=<X_1, W_{(2t)^\nu}, Z_{2t}>, \nu\not=-\frac{1}{2}, a=-2\nu$,

$\mathfrak{L}_{3,9}=<X_1, W_{1/\sqrt{2t}}, Z_{2t}+4p \alpha X_{\sqrt{2t}}>$, $\forall \alpha\in \mathbb{R}, a=1$.

$\mathfrak{L}_{3, 14}=<\beta X_1+Y_1, X_{{(2t)}^\frac{1-a}{2}}+\epsilon\beta Y_{{(2t)}^\frac{1-a}{2}}, Z_{2t}>$, $\forall a\not=1, \epsilon \beta^2\not=1$.

$\mathfrak{L}_{3, 15}=<\beta X_1+Y_1, X_{{(2t)}^\frac{1-a}{2}}+\epsilon\beta Y_{{(2t)}^\frac{1-a}{2}}, Z_{2t}+\beta X_g+Y_g>, \forall a\not=1,\epsilon \beta^2=1$, $\forall g\in C^\infty(\mathbb{R})$.

$\mathfrak{L}_{3, 16}=<X_1,Y_1, Z_{2t}>$, $a=1$.

$\mathfrak{L}_{3, 27}=<W_{t^{-\nu}}, \beta X_1+Y_1, Z_{\nu^{-1}t}>$, $\forall \beta,\nu\in R$, $a=(2\nu)^{-1}\not=0$.

$\mathfrak{L}_{3, 30}=<W_{e^{-t}}, W_{e^{-dt}}, Z_1>$, $\forall a=d\in \mathbb{R}$.

5. Complex action on ideal with $M=\left[\begin{array}{cc}
   a& 1\\
    -1& a\\
\end{array}\right], \forall a\in \mathbb{R}:$

$\mathfrak{L}_{3, 17}$ =$\{A_1, A_2, A_3\}$, where $A_1=X_{\phi}+Y_\Psi,$ $ A_2=X_\psi+Y_\Phi$ and $A_3=Z_1+X_f+X_g$; $\phi=e^{-a t}\cos t, \psi=e^{-a t}\sin t$, $\Phi=\rho \phi+\beta \psi, \Psi=\beta \phi-\rho \psi$ and $\{f, g\}$ are given by (\ref{fg1}) and $\epsilon=\rho^2+\beta^2, \forall \delta, \sigma, \beta\in \mathbb{R}$ and $\rho\in \mathbb{R}\backslash\{0\}$.

$\mathfrak{L}_{3, 18}=\{A_1, A_2, A_3\}$ where $A_1=X_{\phi}+Y_\Psi,$ $ A_2=X_\psi+Y_\Phi$ and $A_3=Z_1+X_f+X_g$; $\phi=e^{-a t}\cos t, \psi=e^{-a t}\sin t$, $\Phi=\beta \psi, \Psi=\beta \phi$ and $f$ and $g$ are given by (\ref{fg2}) $\forall f\in C^\infty(\mathbb{R})$ and $\epsilon=\beta^2, \forall \beta\in \mathbb{R}$.

$\mathfrak{L}_{3, 31}=<W_{e^{-at}\cos t}, W_{e^{-at}\sin t}, Z_{1}>$, $\forall a\in \mathbb{R}$.

6. Jordan action on idea  with $M=\left[\begin{array}{cc}
    1& 0\\
    1& 1\\
\end{array}\right]:$

$\mathfrak{L}_{3,21}=<\beta X_1+Y_1, X_{-\frac{\beta}{2} \ln t+\nu}+Y_{-\frac{1}{2}\ln t+\sigma}, Z_{2t}>$, $\forall \nu, \sigma\in \mathbb{R}$  and $\epsilon\beta^2=1$.

$\mathfrak{L}_{3, 22}=<\beta X_1+Y_1, X_{-\frac{\beta}{2} \ln t}+Y_{-\frac{1}{2}\ln t}, Z_{2t}+\beta X_g+Y_g>$, $\forall g\in C^\infty(\mathbb{R})$ and $\epsilon\beta^2=1$.

$\mathfrak{L}_{3, 32}=<W_{e^{t}}, W_{te^{t}}, Z_{-1}>$.

7. The simple Lie algebra sl$(2,\mathbb{R})$:

$\mathfrak{L}_s=<Z_1, Z_t, Z_{t^2}>$ [see (\ref{sp})].

\section{Conclusions}
The main results of this paper are summed up in Theorems \ref{thh4}-\ref{thh6} providing representatives of the conjugacy classes of one-, two-, and three- dimensional real subalgebras of the 2D-CNLS algebra. The classification is performed  under the action of 2D-CNLS group on 2D-CNLS algebra. We obtain several inequivalent classes for subalgebras with different dimensions:  four classes of one-dimensional subalgebras, twelve classes of two- dimensional subalgebras and thirty three classes of three- dimensional subalgebras. The classification shows the finite dimensional algebra structure of infinite dimensional Lie algebra of the 2D-CNLS system (\ref{e1}). It will provide the way to know the infinite number of the invariant solutions of the 2D-CNLS equations under the 2D-CNLS group with finite solutions of the same equations. To construct invariant solutions to system (\ref{e1}), two equivalent approaches can be adopted. One is to perform the symmetry reduction using a general element $V$ expressed by (\ref{V}). Another procedure is to make use of the classification of subalgebras of the 2D-CNLS algebra, shown in previous sections. In both cases, we have standard procedure to get the invariant solutions\cite{B1,B2}. In the second method, we first go through the procedure using the representatives of each congjugacy classes and then use the transformations (\ref{e19}) and (\ref{e17}) to produce the solutions belong to the equivalent classes of the representatives. The number of the representatives shows the finite number of inequivalent classes of the invariant solutions of the 2D-CNLS system (\ref{e1}). Moreover, we have noted that there are four arbitrary functions in the obtained solutions. This may be useful to solve initial or initial- boundary problems of the system (\ref{e1}) by the renormal group methods\cite{I3}. Hence, the state is now set for performing the actual symmetry reduction of the 2D-CNLS system (\ref{e1}) and obtaining \textit{all} solutions of the 2D-CNLS equations that are invariant under the action of 2D-CNLS subgroups. For this rich topic, due to the lack of space and different focus, we will discuss it in detail in another article.

{\bf Acknowledgments}
This work is supported by National Natural Science Foundation of China (No. 11571008).

\end{document}